\documentclass[iop,appendixfloats]{emulateapj}

\shorttitle{LG monitoring. IV}
\shortauthors{Navabi et al.}
\usepackage[usenames,dvipsnames]{color}
\usepackage{amsmath}
\usepackage{float}
\usepackage{ctable} 
\usepackage{graphicx}
\usepackage{natbib}
\begin{document}
                                
\title{The Isaac Newton Telescope monitoring survey of Local Group dwarf galaxies -- IV. The star formation history of Andromeda VII derived from long period variable stars}

\author{Mahdieh Navabi\altaffilmark{1},
Elham Saremi\altaffilmark{1}, 
Atefeh Javadi\altaffilmark{2,1}, 
Majedeh Noori\altaffilmark{3}, 
Jacco~Th.~van~Loon\altaffilmark{4},
Habib G. Khosroshahi\altaffilmark{1,5},
Iain McDonald\altaffilmark{6,7},
Mina Alizadeh\altaffilmark{3},
Arash Danesh\altaffilmark{5},
Ghassem Goz­aliasl\altaffilmark{8,9,10}
Alireza Molaeinezhad\altaffilmark{11,12},
Tahere Parto\altaffilmark{1,13},
Mojtaba Raouf\altaffilmark{14}}

\altaffiltext{1}{School of Astronomy, Institute for Research in Fundamental Sciences (IPM), P.O.~Box 1956836613, Tehran, Iran, mahdieh.navabi@ipm.ir}
\altaffiltext{2}{Department of Physics, Sharif University of Technology, P.O.\ Box 11155-9161, Tehran, Iran}  
\altaffiltext{3}{Department of Physics, University of Zanjan, University Blvd., P.O.\ Box: 45371-38791, Zanjan, Iran}
\altaffiltext{4}{Lennard-Jones Laboratories, Keele University, ST5 5BG, UK}
\altaffiltext{5}{Iranian National Observatory, Institute for Research in Fundamental Sciences (IPM), Tehran, Iran}
\altaffiltext{6}{Jodrell Bank Centre for Astrophysics, Alan Turing Building, University of Manchester, M13 9PL, UK}
\altaffiltext{7}{Department of Physical Sciences, The Open University, Walton Hall, Milton Keynes, MK7 6AA, UK}
\altaffiltext{8}{Finnish Centre for Astronomy with ESO (FINCA), Quantum,University of Turku, Vesilinnantie 5, 20014 Turku, Finland}
\altaffiltext{9}{Department of Physics, University of Helsinki, P.O.~Box 64, 00014 Helsinki, Finland}
\altaffiltext{10}{Helsinki Institute of Physics, University of Helsinki, P.O.~Box 64, 00014 Helsinki, Finland}
\altaffiltext{11}{ Department of Physics, University of Oxford, Keble Road, OX1 3RH Oxford, UK}  
\altaffiltext{12}{Institute of Astronomy, University of Cambridge, Madingley Road, Cambridge CB3 0HA, UK}
\altaffiltext{13}{Physics Department, Alzahra University, Vanak, 1993891176, Tehran, Iran}
\altaffiltext{14}{Korea Astronomy and Space Science Institute, 776 Daedeokdae-ro, Yuseong-gu, Daejeon 34055, Republic of Korea}

%
\begin{abstract}

We have examined the star formation history (SFH) of Andromeda VII (And\,VII), the brightest and most massive dwarf spheroidal (dSph) satellite of the Andromeda galaxy (M\,31). Although M\,31 is surrounded by several dSph companions with old stellar populations and low metallicity, it has a metal-rich stellar halo with an age of 6--8 Gyr. This indicates that any evolutionary association between the stellar halo of M\,31 and its dSph system is frail. Therefore, the question is whether And\,VII (a high-metallicity dSph located $\sim$220 kpc from M\,31), can be associated with M\,31's young, metal-rich halo. Here, we perform the first reconstruction of the SFH of And\,VII employing long-period variable (LPV) stars. As the most-evolved asymptotic giant branch (AGB) and red supergiant (RSG) stars, the birth mass of LPVs can be determined by connecting their near-infrared photometry to theoretical evolutionary tracks. We found 55 LPV candidates within two half-light radii, using multi-epoch imaging with the {\it Isaac Newton} Telescope in the $i$ and $V$ bands. Based on their birth mass function, the star-formation rate (SFR) of And\,VII was obtained as a function of cosmic time. The main epoch of star formation occurred $\simeq 6.2$ Gyr ago with a SFR of $0.006\pm0.002$ M$_\odot$ yr$^{-1}$. Over the past 6 Gyr, we find slow star formation, which continued until 500 Myr ago with a SFR $\sim0.0005\pm0.0002$ M$_\odot$ yr$^{-1}$. We determined And\,VII's stellar mass $M=(13.3\pm5.3)\times10^6$ M$_\odot$ within a half-light radius $r_{\frac{1}{2}}=3.8\pm0.3$ arcmin and metallicity $Z=0.0007$, and also derived its distance modulus of $\mu=24.38$ mag.

\end{abstract}

%
\section{Introduction}
\label{introduction}

Dwarf spheroidal (dSph) galaxies are early-type galaxies devoid of gas and dust. As most dSphs lie in the halos of their massive host galaxies, their gas has been stripped through interactions with their host galaxies and thus their star formation has halted long ago \citep{2006A&A...447..473B,2009ApJ...696..385G,2012ApJ...752...45T}. Since there is no sign of gaseous content and young stars in dSphs, their earlier star formation history (SFH) plays a decisive role in the scenario of formation and evolution of these low-mass galaxies. The time and speed of quenching for a dwarf can be determined from its SFH, and used to quantify the effects of halo mass, infall time and vicinity to a host, as well as stellar feedback and heating of gas at the epoch of reionization.

To explore the link between host galaxies and their dSph companions, we here study Andromeda VII, the most massive and extended dSph associated with M\,31. And\,VII (also known as the Cassiopeia dwarf) was discovered by Karachentsev and Karachentseva in \citeyear{1999A&A...352..363K} as a satellite of M\,31, along with two companions, And\,V and And\,VI. This satellite is located near the Galactic plane ($l=109.5~{\deg}$, $b=-10~\deg$) and exhibits foreground reddening of $E(B-V)=0.194$ mag \citep{2006MNRAS.365.1263M}. The distance to And\,VII has been determined from the tip of the red giant branch (RGB) as $\mu=24.41$ mag (763 kpc) \citep{2005MNRAS.356..979M}. From scaling the color--magnitude diagram (CMD) a distance modulus of $\mu=24.5$ mag was estimated by \cite{2003A&A...408..111K} and $\mu=24.58$ mag by \cite{2014ApJ...789..147W}.

And\,VII contains the largest stellar mass between all M\,31 dSph satellites ($M_*=19.73\times10^6$ M$_\odot$) and it is the eighth most massive galaxy in the entire M\,31--M\,33 group \citep{2018ApJ...868...55M}. Accordingly, it is one of the most luminous satellites of M\,31 with $M_V=-13.3\pm0.3$ mag and $L_V=1.8\times10^7$ L$_\odot$ \citep{2010ApJ...711..671K}. Besides, it has a higher metallicity than most of M\,31's dSphs ([Fe/H] $=-1.4\pm0.3$ obtained by \cite{1999ApJ...511L.101G} from a Keck/LRIS photometric study and [Fe/H] $\sim-1.3\pm0.1$ by \cite{2014ApJ...790...73V,2020AJ....159...46K,2020ApJ...895...78W} from spectra).

Despite the present large separation of And\,VII from M\,31 ($\sim220$ kpc), this dwarf is one of the most $\alpha$-element-enhanced systems, showing a flat trend in [$\alpha$/Fe] with [Fe/H]. This would suggest a closer proximity of And\,VII to M\,31 in the past, which purportedly caused it to cease star formation \citep{ 2014ApJ...790...73V}. Furthermore, the [$\alpha$/Fe] pattern of M\,31's outer halo is consistent with that of And\,VII \citep{2020AJ....160...41G}. The other feature that supports an ancient, close encounter is the equal radial velocity of And\,VII and the peak of M\,31's halo velocity \citep[$v_{\rm rad}=-309.4\pm2.3$ km s$^{-1}$;][]{2012ApJ...752...45T}.

To probe the star formation and enrichment histories in dwarf galaxies, CMD synthesis has been the method of choice in recent times \citep[e.g.][]{1997NewA....2..397D,2007ApJ...659L..17C,2014ApJ...789..147W,2017ApJ...837..102S,
	2018MNRAS.479.5035D}. Despite the popularity of the CMD method, this method can not be a suitable approach for all galaxies. For example, \cite{2014ApJ...789..147W} derived the SFH of 40 Local Group dwarf galaxies based on CMD analysis, but they admitted that they were not successful in determining the anomalous SFH of And\,VII and Cetus dwarfs, because their photometry was too shallow, and key CMD features used to derive the SFH were missing (e.g., horizontal branch or red clump). In this study, we employ an alternative method based on long-period variable (LPV) stars to reconstruct the SFH of And\,VII. This SFH method was described by \cite{2011MNRAS.414.3394J} and has been successfully applied to M\,33 and several dwarfs \citep{2017MNRAS.464.2103J,2014MNRAS.445.2214R,2017MNRAS.466.1764H,
	2019MNRAS.483.4751H,2019IAUS..344..125S}.

LPV stars are mostly asymptotic giant branch (AGB) stars that reach the highest luminosity at their final stages of evolution, and hence there is a direct relationship between their luminosity and birth mass. Therefore, they enable us to derive the SFH across a broad range of ages, from $\sim30$ Myr to $\sim10$ Gyr \citep{2008A&A...482..883M}. In addition, to trace the more recent SFH $\sim10$--30 Myr ago, red supergiant (RSG) stars with higher mass ($\sim8$--30 M$_\odot$) can be used in a similar manner \citep{2005AAS...20718213L,2010ASPC..425..103L}. The cool evolved AGB stars ($\sim0.8$--8 M$_\odot$) reach high luminosity ($\approx 1000$--60\,000 L$_\odot$) and low temperature ($T\approx3000$ K) towards the end of their lifetime \citep{2018A&ARv..26....1H,2018AJ....156..112Y}. The corresponding low surface gravity leads to strong radial pulsations which, in turn, facilitates the production of dust and a strong stellar wind, replenishing the interstellar medium \citep[ISM;][]{2019A&A...626A.100B}. LPV stars are easily detected and identified in nearby galaxies due to their pulsation on timescales of months to years \citep{2019IAUS..339..336S}. At the peak of their spectral energy distribution, they stand out most conspicuously at near-infrared (near-IR) wavelengths, which helps diminish the effect of dust extinction and reddening \citep{1983ARA&A..21..271I,2007ApJ...657..810D}.

We conducted an optical survey of nearby galaxies (the most complete sample so far) with the 2.5-m Isaac Newton Telescope (INT) over nine epochs \citep{2020ApJ...894..135S}. Our main objectives include: identify all LPVs in the dwarf galaxies of the LG accessible in the Northern hemisphere, and then determine the SFHs from their luminosity distribution; obtain accurate time-averaged photometry for all LPVs; obtain the pulsation amplitude of them; determine their radius variations; model their spectral energy distributions (SEDs); and study their mass loss as a function of stellar properties such as mass, luminosity, metallicity, and pulsation amplitude. This is Paper\,IV in the series, with identifying the LPV stars in And\,VII and estimating of the SFH of this galaxy. The article layout is as follows: the description of the data is given in Sect.~\ref{sec:observations}. The data reduction and a brief review of the photometry methods are presented in Sect.~\ref{sec:ANALYSIS}. The catalog is described in Sect.~\ref{sec:catalog}, followed by the method for finding the LPVs \ref{sec:variable stars}. In Sect.~\ref{Color-Magnitude Diagram}, we discuss And\,VII's CMD and obtain the AGB and RGB tips and the distance modulus. The method of deriving the SFH is described in Sect.~\ref{sec:sfh}. The results and conclusions are explained in Sect.~\ref{sec:result} and~\ref{sec:conclusions}, respectively.

\begin{figure*}
	\centering
	\includegraphics[width=\linewidth, height=1.1\columnwidth]{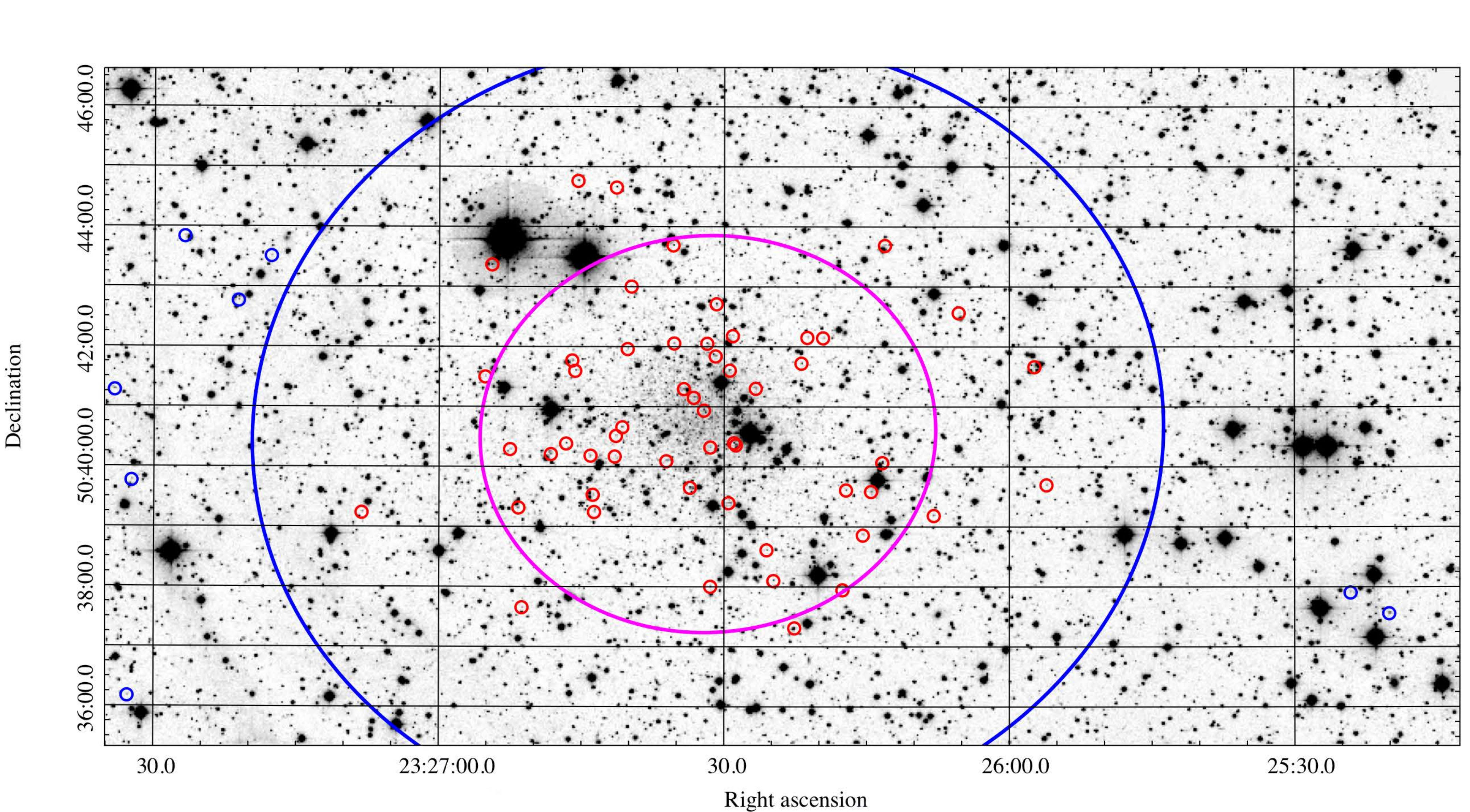}
	\caption{The master WFC image of the And\,VII dSph galaxy along with the spatial location of LPV candidates (inside $2r_{\frac{1}{2}}$ in red and outside $2r_{\frac{1}{2}}$ in blue). The half-light radius is marked with a magenta ellipse $r_{\frac{1}{2}}=3\rlap{.}^\prime8$ with an ellipticity of 0.13.  The $2r_{\frac{1}{2}}$ radius is displayed in blue.}
	\label{fig:lpvs}
\end{figure*}

\section{Observations}
\label{sec:observations}

The data set was obtained with WFC/INT over a period of three years (June 2015 to October 2017) from a survey of the majority of dwarf galaxies in the Local Group, including 43 dSph, six dIrr, six dTrans and four globular clusters, all visible in the northern hemisphere \citep{2017JPhCS.869a2068S,2020ApJ...894..135S}. The WFC is an optical mosaic camera at the prime focus of the 2.5m {\it Isaac Newton} Telescope (INT) in La Palma, Spain. It consists of four $2048\times4096$ CCDs, with a pixel size of $0\rlap{.}^{\prime\prime}33$ pixel$^{-1}$. A combined mosaic of And\,VII from $i$ and $V$ bands is shown in Figure~\ref{fig:lpvs} ($11.26\times22.55$ arcmin$^2$). 

The survey was designed to measure the amplitude and mean brightness of LPV stars, which are variable due to radial pulsations on time-scales from $\approx60$ days \citep{2016ApJ...823L..38M} for low-luminosity AGB stars to $\approx700$ days for intermediate-luminosity AGB stars \citep{1992ApJ...397..552W}, as well as periods up to $\sim2000$ days for RSGs \citep{2006AstL...32..263S}. For this purpose, observations that were spaced a few months apart were taken over eight epochs in the WFC Sloan $i$ filter; except for the first epoch when the WFC RGO $I$ filter was used and subsequently transformed to Sloan $i$ (explained in section~\ref{sec:ANALYSIS}). Additionally, five epochs in the Harris $V$ filter were observed to obtain colour information. The details of the observations of And\,VII are listed in Table~\ref{tab:log}.

\begin{table}
	\centering
	\caption{\label{table:log} Log of WFC observations of the And\,VII dwarf galaxy}
	\label{tab:log}
	\begin{tabular}{cccccc}
		\hline
		\hline
		Date    & Epoch & Filter & $t_{\rm exp}$ & seeing & Airmass \\
		(y m d) &       &        & (sec)       & arcsec & \\
		\hline
		\hline
		2015 06 18 & 1 & I &  555 & 1.87 & 1.339 \\
		2015 06 18 & 1 & R &  735 & 1.89 & 1.262 \\
		2015 06 18 & 1 & V & 1200 & 1.98 & 1.174 \\
		2016 02 10 & 1 & i &  629 & 2.17 & 1.821 \\
		2016 06 13 & 2 & i &  796 & 3.22 & 2.259 \\
		2016 08 10 & 3 & i &  555 & 2.24 & 1.228 \\
		2016 08 12 & 3 & i &  683 & 2.13 & 1.332 \\
		2016 08 12 & 2 & V &  126 & 2.05 & 1.084 \\
		2016 10 20 & 4 & i &  555 & 2.25 & 1.102 \\
		2017 01 30 & 5 & i &  556 & 1.98 & 1.593 \\
		2017 08 01 & 6 & i &  555 & 1.97 & 1.803 \\
		2017 08 01 & 3 & V &  736 & 2.08 & 1.576 \\
		2017 09 02 & 7 & i &  556 & 2.08 & 1.089 \\
		2017 09 02 & 4 & V &  736 & 2.15 & 1.079 \\
		2017 10 06 & 8 & i &  555 & 2.09 & 1.204 \\
		2017 10 08 & 5 & V &  736 & 2.40 & 1.221 \\
		\hline
		\hline
	\end{tabular}
	\vspace{1ex}
	
	{\raggedright Note: the observation in the $R$ band is used for transforming RGO $I$ to Sloan $i$ band.\par}
\end{table}
\section{Data analysis}
\label{sec:ANALYSIS}

Before photometry, the raw CCD frames must be processed. To this end, we used {\sc theli} (Transforming HEavenly Light into Image) as an image processing pipeline with astronomical software that is adapted to multi-chip cameras for the data reduction process \citep{2020ApJ...894..135S}. Point Spread Function (PSF) photometry calibration was performed with the {\sc daophotII} package \citep{1987PASP...99..191S} using the {\sc daophot/allstar/allframe} routines. Further details, including data processing, photometric calibration, and relative calibration are given in \cite{2020ApJ...894..135S}.

For obtaining the epochs, we used Sloan-$i$ and Harris $V$ filters; except for the first night (18 June 2015) that had another filter (the WFC RGO $I$ filter). Hence, we used transformation equations extracted from \cite{2006A&A...460..339J} to estimate magnitudes for all stars in the Sloan $i$ filter. The $R$-band image of 18 June 2015 was used to transform the $I$-band data into Sloan-$i$ magnitudes using the following relation:
\begin{equation}
i-I = (0.251 \pm 0.003)\times (R-I) + (0.325 \pm {0.002}) 
\label{tfr}
\end{equation}
Before performing this calculation, it was necessary to determine the zero point of each frame to provide the {\sc allframe} lists in unison. Also, $R$  and $I$ {\sc allframe} lists were used to select program stars in the {\sc daogrow} and {\sc collect} routines to correct the aperture photometry of Sloan-$i$ magnitudes, but they have not been used in the {\sc newtrial} section. Hereafter, the stellar magnitudes from all frames are related to two bands (Sloan $i$ and $V$) to make the master mosaic of And\,VII in the final stage of {\sc newtrial}.

\subsection{Photometry evaluation}
\label{sec:Evaluation of Photometry}

\begin{figure}
	\centering
	\includegraphics[width=\columnwidth,height=0.67\columnwidth]{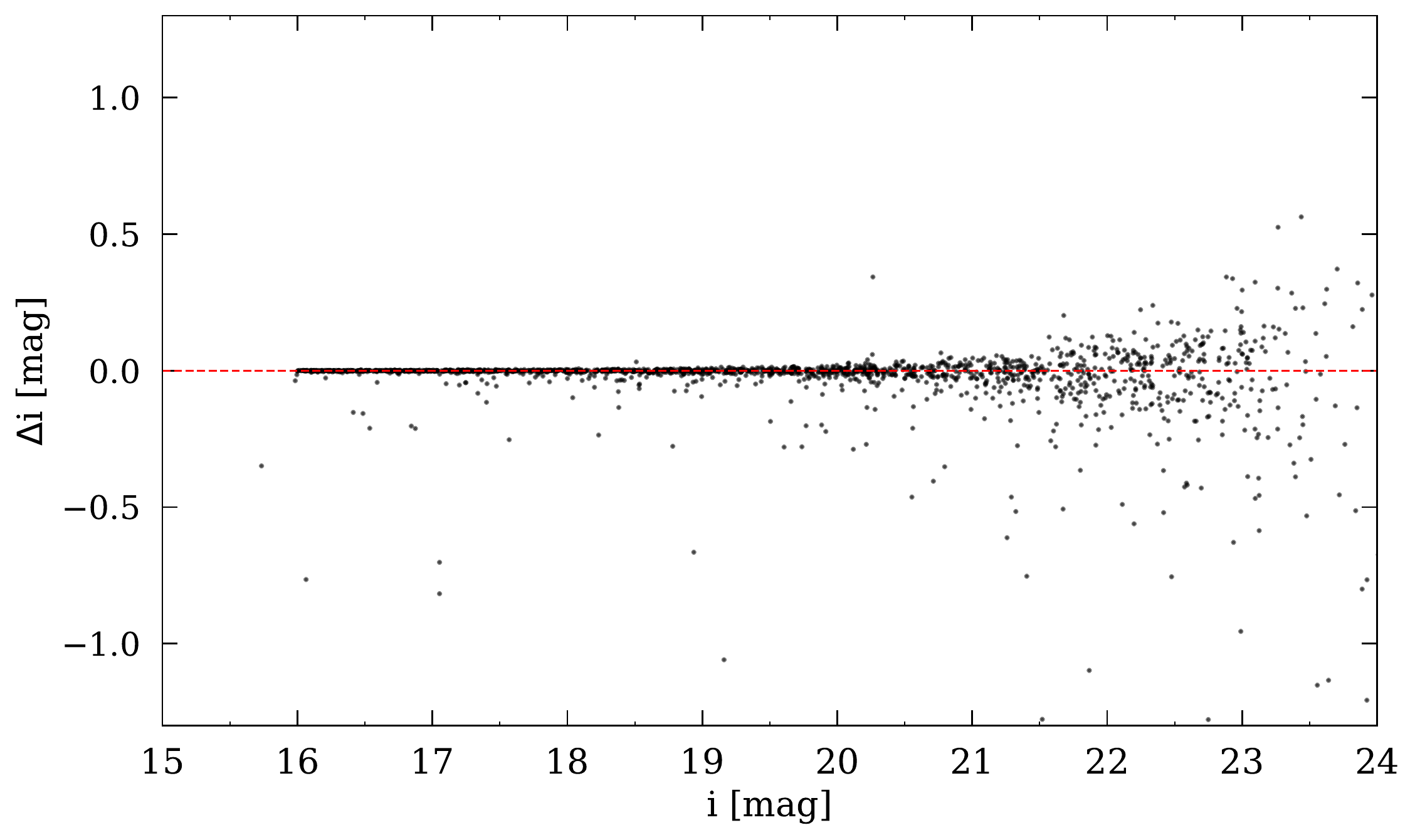}
	\caption{The scatter of difference in magnitudes between the input and the recovered artificial stars.}
	\label{fig:scatter}
\end{figure}

The photometric completeness was tested with the {\sc addstar} task in the {\sc daophot} package; $2500$ artificial stars were added to the master mosaic in the $i$ and $V$ bands. These were split over five trials, to avoid increasing the crowding level. The positions of stars were selected randomly in the images, then Poisson noise was added to them. We specified the magnitude of artificial stars between 16--25 mag in one-mag bins (similar bins for $V$ band). Then, the photometric reduction was repeated to assess the precision of photometry. Figure~\ref{fig:scatter} shows the difference in magnitudes between the input and the recovered artificial stars using the {\sc daomaster} routine. The magnitude difference in $i$-band is very small, $|\Delta i| < 0.1$ mag up to $i\approx 22$ mag.

\begin{figure}
	\includegraphics[width=\columnwidth]{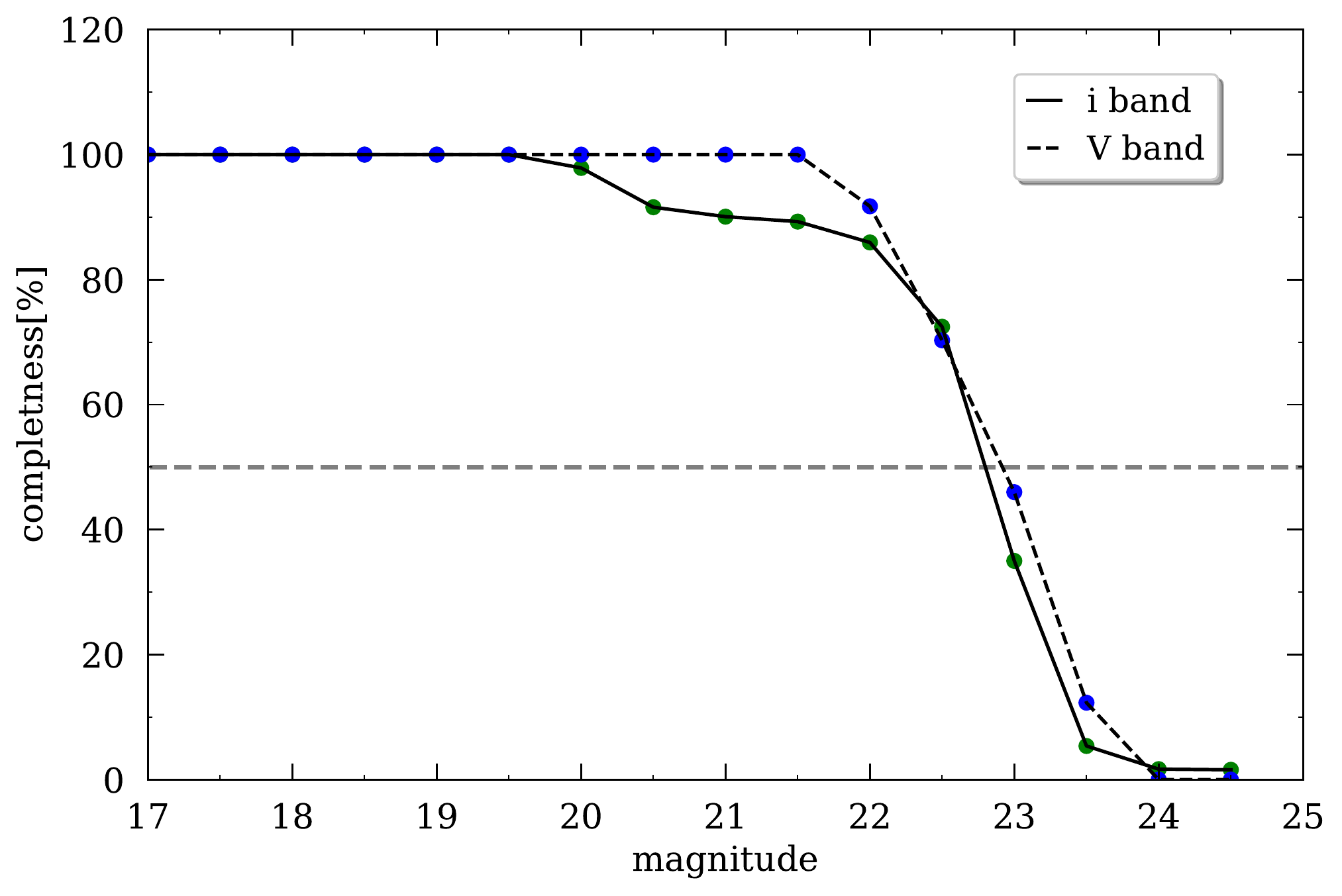}
	\caption{Completeness as a function of $i$-band (solid line) and $V$-band (dashed line) magnitude.}
	\label{fig:completeness}
\end{figure}

Also, we found the fraction of recovered stars in the 0.5-mag bins in both bands to determine the completeness. Figure~\ref{fig:completeness} displays the percent of the recovered stars, which decreases with increasing magnitudes and reaches a completeness of 50\% at $i=22.8$ mag and $V=23$ mag. The important range of magnitudes in this project is between the RGB tip ($i=21.3$ mag, $V=22.35$ mag) and the AGB tip ($i=17.63$ mag, $V=17.81$ mag) (see Section~6).

We have performed the photometric calibration process in three steps as explained by \citet[][see also Section ~\ref{sec:Callibration}]{2020ApJ...894..135S}, and obtained the final catalog of 14\,092 objects with the {\sc newtrial} routine. 


\subsection{Calibration}
\label{sec:Callibration}

We have performed the photometric calibration process in three steps:
\begin{enumerate}
	\item{Aperture correction:} 
	With the use of the {\sc daogrow} and {\sc collect} routines, the differences in magnitude between the PSF-fitting and large-aperture photometry of the program stars ($20-30$ isolated bright stars in each frame) were calculated \citep{1990PASP..102..932S}. Finally, {\sc newtrial} applied these aperture corrections to all stars for each frame.
	
	\item{Transformation to the standard photometric system:}
	The transformation equation for each frame is constructed based on zero point and atmospheric extinction. Also, for frames without a standard-field observation, we used the mean of other zero points. The {\sc ccdave} routine applied the transformation equation on the program stars for each frame. Then, program stars as local standards were used in the {\sc newtrial} routine to calibrate all the other stars.
	
	\item{Relative calibration:} 
	Relative photometry achieves a balance between all epochs. Thus, similar magnitudes are obtained for a non-variable star in all epochs. The calculated relative deviation values for frames are between 0.6--9 mmag. The relative deviation of each frame was subtracted from the magnitudes of all stars in that frame. This correctly separates variable from non- variable sources \citep{2020ApJ...894..135S}.
\end{enumerate}

\section{Properties of the dataset}
\label{sec:catalog}

\subsection{Determining the half-light radius of And\,VII}
\label{sec:rh}

\begin{figure}
	\includegraphics[width=\columnwidth,height=0.97\columnwidth]{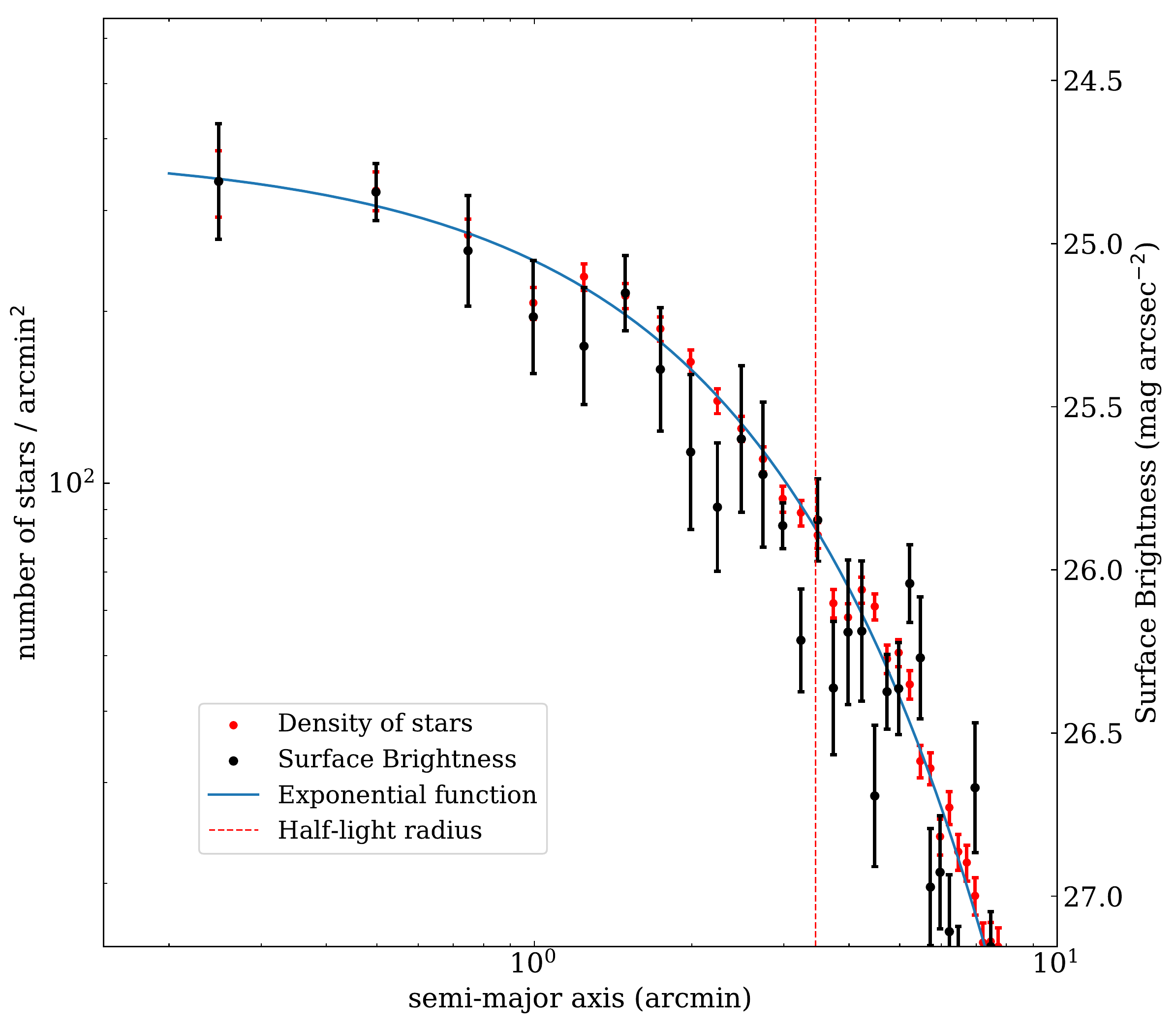}
	\caption{The log--log plot of the stellar number density of And\,VII in red points, and the surface brightness in black points on the right vertical axis. The blue solid curve depicts the best exponential fit to the data. Error bars are calculated from the Poisson uncertainty in the counts. We determined $r_{\frac{1}{2}}=3\rlap{.}^\prime8\pm0\rlap{.}^\prime3$.}
	\label{fig:SB}
\end{figure}

Figure~\ref{fig:SB} shows And VII's surface brightness and stellar number density on concentric elliptical annuli as a function of the semi-major axis in black and red points. The best-fitting exponential profile is overlaid in blue, using a least-squares minimization technique. The error bars are derived from the Poisson error on the photon counts. To obtain And\,VII's half-light radius, $r_{\frac{1}{2}} $, the total extrapolated luminosity of the exponential profile is calculated by integrating up to $r\rightarrow\infty$:
\begin{equation}
L(r) = 2\pi\epsilon\int_0^r r\,S(r)\,dr
\end{equation}
where $S(r)$ is the radial profile with an exponential law, and  $\epsilon=0.13$ is the measured ellipticity of the galaxy \citep{2006MNRAS.365.1263M}. We thus obtained the semi-major axis of the half-light ellipse, $r_{\frac{1}{2}}=3\rlap{.}^\prime8\pm0\rlap{.}^\prime3$. This agrees with previous values of $3\rlap{.}^\prime5\pm0\rlap{.}^\prime1$ \citep{2006MNRAS.365.1263M} and $791\pm45$ pc corresponding to $3\rlap{.}^\prime56\pm0\rlap{.}^\prime20$ \citep{2010ApJ...711..671K}.

\begin{figure}
	\includegraphics[width=\columnwidth]{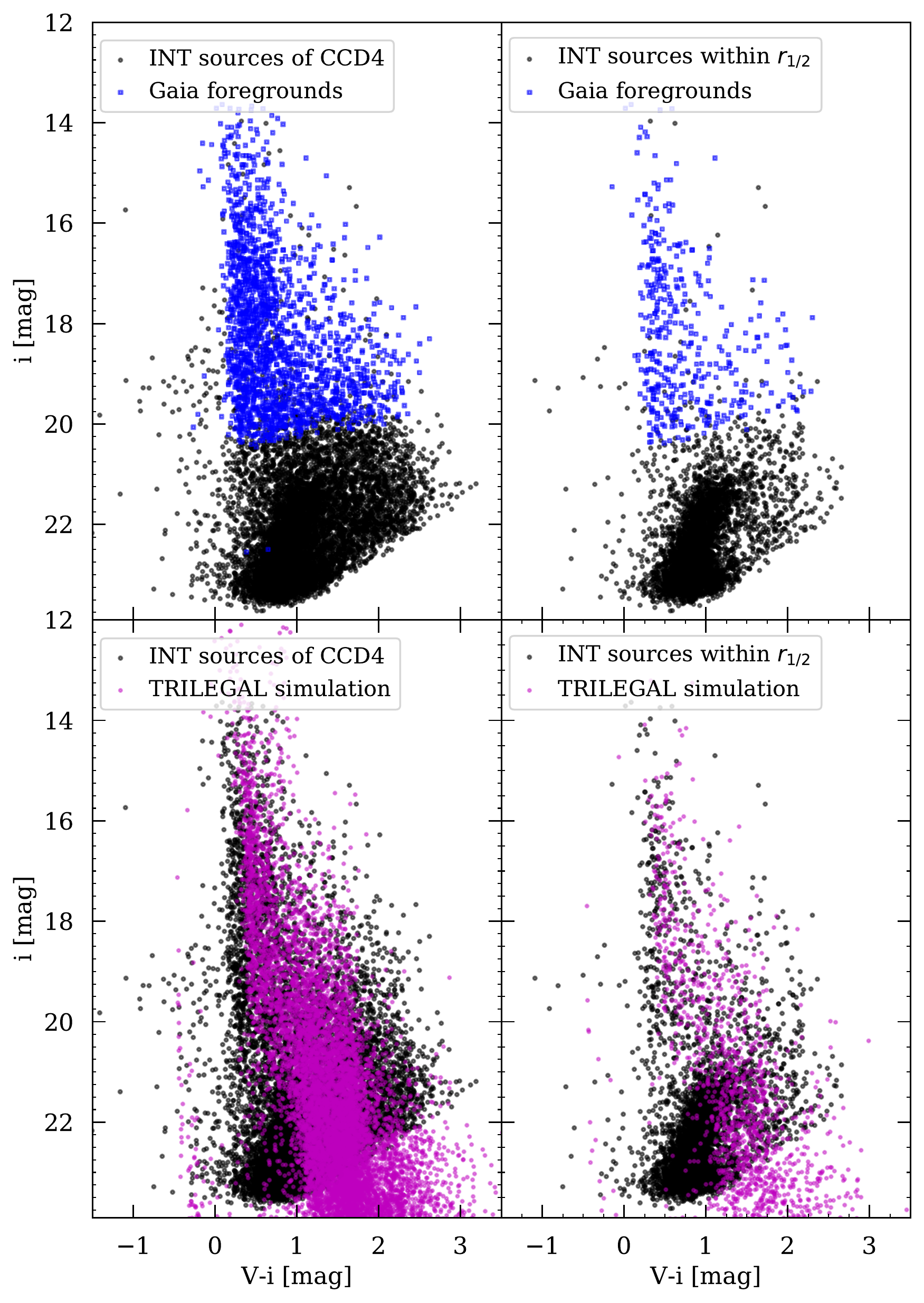}
	\caption{The Gaia foreground stars (in blue) and predicted contamination of foreground stars (in magenta) as simulated with {\sc trilegal} \citep{2005A&A...436..895G} are presented in two panels: the left concerns a 0.07 deg$^2$ field while the right panel is restricted to 0.01 deg$^2$ centered on And\,VII corresponding to the half-light radius of the galaxy. INT sources are in black.}
	\label{fig:foreground}
\end{figure}

\subsection{Foreground contamination}
\label{Sky Foreground}

And\,VII is located at low Galactic latitude and is contaminated by foreground stars of the Milky Way, hence disposing of these foreground stars is essential for accurate recognition of LPV stars in the dwarf galaxy. Therefore, we cross-matched our catalog with the second Gaia data release \citep[Gaia DR2;][]{2018A&A...616A...1G} to remove the contamination of foreground stars in this direction. To enhance the accuracy of foreground star selection, we applied particularly strict criteria on the proper motion and parallax of Gaia stars, which are similar to the criteria of \cite{2020ApJ...894..135S}. Accordingly, 1872 stars were found as foreground stars but only 380 of them were located within $r_{\frac{1}{2}}$ from the center of And\,VII. The top-left panel of Figure~\ref{fig:foreground} shows the Gaia foreground stars in blue and And\,VII's candidate members in black for the CCD 4 field. The right panel is similar to the left panel but for stars located within $r_{\frac{1}{2}}$. The completeness limit of Gaia is near $G\sim17$ mag and unfortunately stars with $i>20.5$ mag are not very well distinguished in this catalog \citep{2018A&A...616A...1G}.

Hence, we also assess the foreground contamination in the direction of And\,VII, and the extinction in the $V$-band of $A_V=0.532$ mag, using the {\sc trilegal} simulation, which reproduces the stellar population artificially \citep{2005A&A...436..895G}. In Figure~\ref{fig:foreground}, the magenta points represent the predicted contamination of two different fields: 0.07 deg$^2$ (about the size of the entire CCD 4 of WFC; left panel) and 0.01 deg$^2$ (the half-light radius of And\,VII; right panel).
We compare the number of foreground contaminants with $i<20.5$ mag obtained via two procedures to estimate the accuracy of {\sc TRILEGAL} prediction. Within $r_{\frac{1}{2}}$, the cross-correlation between our results and Gaia catalog suggested the number of 380 foreground stars that is consistent with the estimation of {\sc TRILEGAL} simulation (392 stars). In the whole region of CCD 4, the number of 1872 foreground stars are found from Gaia catalog, while 2664 foreground stars estimated from {\sc TRILEGAL} simulation. However, the $i=20.5$ mag is not an exact limit in  Figure~\ref{fig:foreground} and with decreasing it to $i=20$ mag, the number of contaminations is reduced to 2152 sources by {\sc TRILEGAL} and 1763 by Gaia. Therefore, for $i>20.5$ mag,  we can benefit {\sc TRILEGAL} simulation and be more sure of estimation in inner regions, which are more important in this study, than outer.

Clearly, the region around the galaxy is extremely contaminated by foreground stars, while the CMD of the inner region ($r<r_{\frac{1}{2}}$) suffers much less contamination. The measured stellar surface densities within $r_{\frac{1}{2}}$ and across the rest of the field are 0.04 and 0.01 arcsec$^{-2}$, respectively. The {\sc trilegal} model estimates Milky Way foreground densities over these two regions of $\sim0.01$ and $\sim0.02$ arcsec$^{-2}$, respectively. Thus, the fraction of contaminating foreground stars is much larger outside $r_{\frac{1}{2}}$ than inside. And\,VII has a large projected distance from M\,31 ($16\rlap{.}{\deg}^2$), hence contamination from M\,31 halo stars is unlikely to be significant \citep{2012ApJ...752...45T}.

\begin{figure}
	\includegraphics[width=\columnwidth,height=\columnwidth]{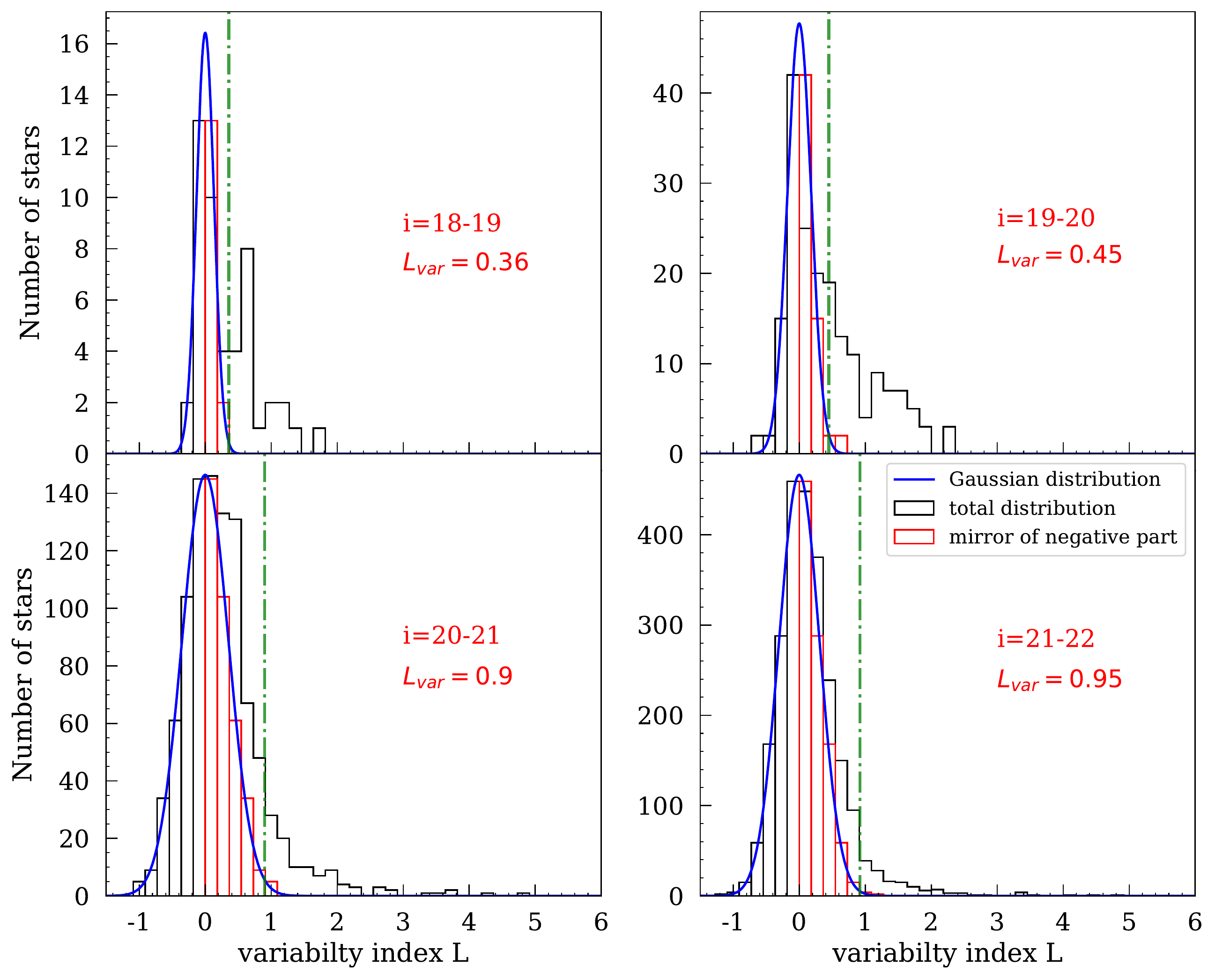}
	\caption{Histograms of the variability index $L$, for several $i$-band magnitude bins. The negative part of each histogram is mirrored (red bins). This is fitted with a Gaussian function to show the expected distribution of non-variable sources. The optimal variability index thresholds are indicated by vertical green-dashed lines.}
	\label{fig:histogram}
	%
	\includegraphics[width=\columnwidth,height=\columnwidth]{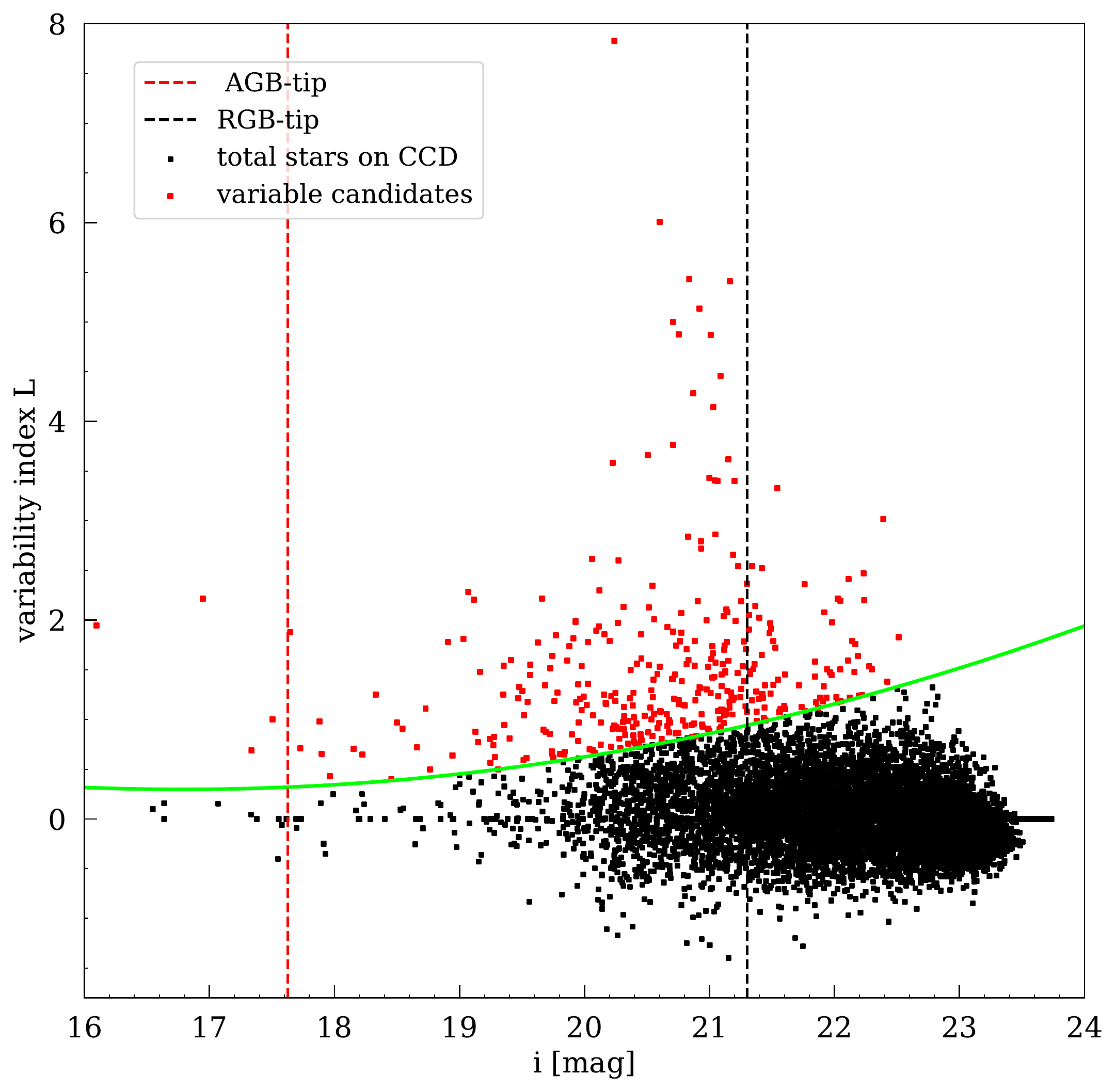}
	\caption{Variability index $L$ of all stars vs.\ $i$-band magnitude. The green, curved line indicates the threshold for different magnitudes. As a reminder, positively identified foreground stars have already been removed.}
	\label{fig:indexl}
\end{figure}

\section{Searching for variable stars}
\label{sec:variable stars}

In order to identify variable stars, we applied a method similar to the {\sc newtrial} program described by \cite{1993AJ....105.1813W} and also developed further by \cite{1996PASP..108..851S}. This method attributes to each star a variability index ($L$) that is derived based on $J$ and $K$ indices. Typically, variable stars show a large positive $J$ index, while the $J$ index tends to zero for stars with random noise. Besides, the $K$ index is described based on the shape of light-curve.
\begin{equation}
L=\frac{J\times K}{0.798}\frac{{\sum_{i=1}^N w_i}}{w_{\rm all}}
\end{equation}
where $\sum w$ is total weight assigned to a star and $w_{\rm all}$ is the total weight that a star would be assigned if observed in all observations. The weight ($w_i$) of each star is 1 in pairs of frames that are close in time compared to the period of variability (expected $\sim100$ days or longer), whilst in a single frame $w_i=0.5$.

After having calculated $L$ for all stars, we determined an optimal variability threshold that separates variables from other stars. The histograms of $L$ are presented in four magnitude ranges surrounding the RGB and AGB tips, in Figure~\ref {fig:histogram}. To determine the variability threshold ($L_{\rm var}$), we mirrored the negative part of each histogram with respect to $L=0$, because we do not expect to have physical variables with $L<0$. Then, a Gaussian function was fitted to these negative parts and their mirrors to resemble the non-variable sources, allowing us to estimate the percentage of variable stars in each bin. The Gaussian fits closely match the histogram at low values of $L$, but deviate from it at higher values, representing the sample of intrinsically variable stars. For each magnitude interval we chose $L_{\rm var}$ beyond which 90\% of objects in a bin are above the Gaussian fit (i.e.\ only 10\% of the subset are expected to be non-variable). The threshold values of $L$ are labeled on Figure~\ref{fig:histogram}.

In Figure~\ref{fig:indexl}, we show the variability index of all stars vs.\ magnitude, along with a polynomial curve fit to the threshold points obtained from histograms. $L_{\rm var}$ increases with $i$-band magnitude, as the photometric uncertainties increase. The red stars with $L>L_{\rm var}$ in the plot are candidate variables if they are not visibly affected by blending. After eliminating these suspect stars, we identified 154 candidate variable stars in the CCD field, most of which are between $i\approx 20$ and 22.5 mag, i.e.\ around the RGB tip.

\begin{figure}
	\includegraphics[width=\linewidth,height=0.85\columnwidth]{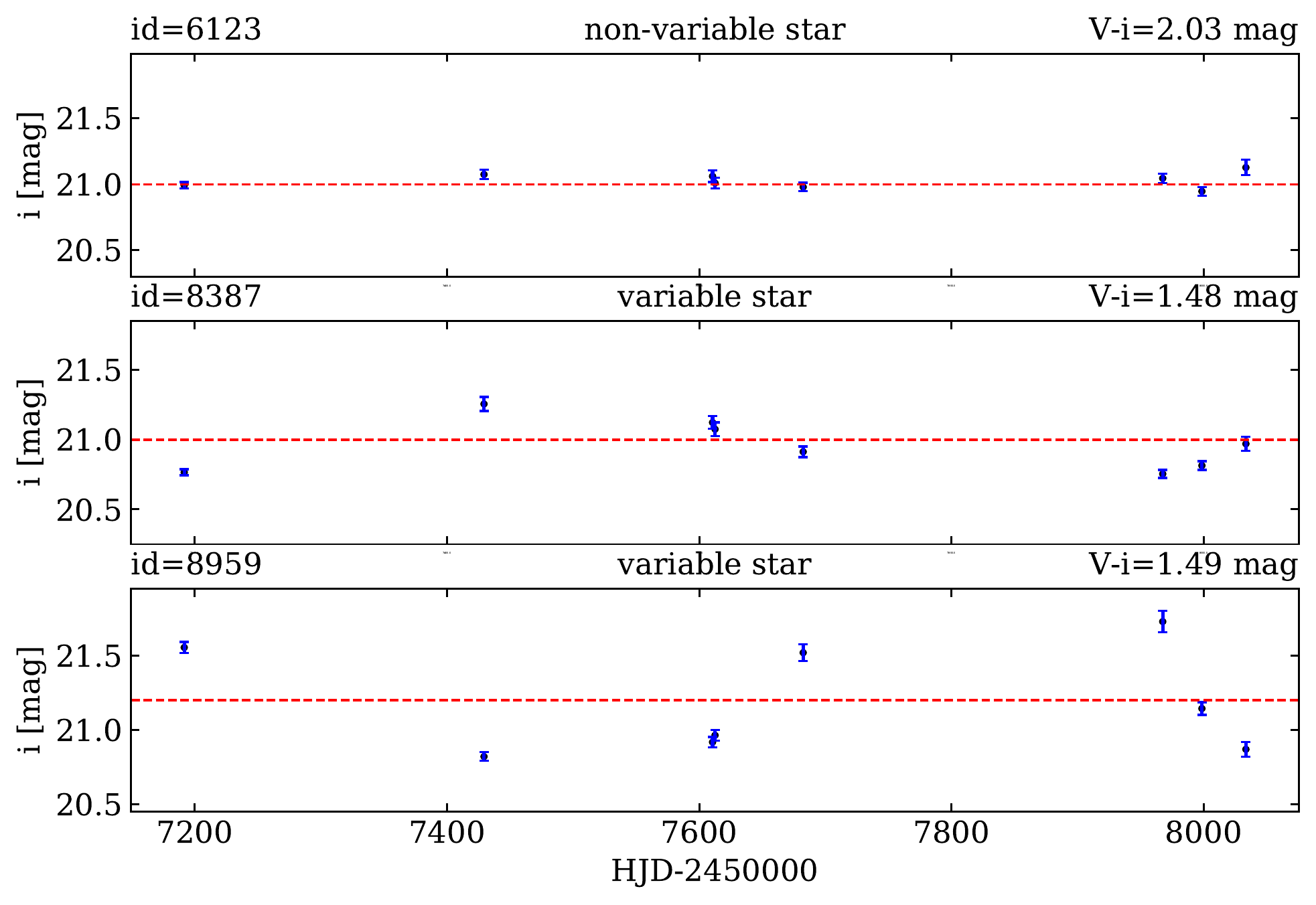}
	\caption{Light-curves of two LPV candidates labelled with green color in Figure~\ref{fig:amplitude}, and a non-variable star in the top panel for comparison.}
	\label{fig:example of variable}
\end{figure}

\subsection{Amplitudes of variability}

\begin{figure}
	\centering
	\includegraphics[width=\columnwidth,height=1.03\columnwidth]{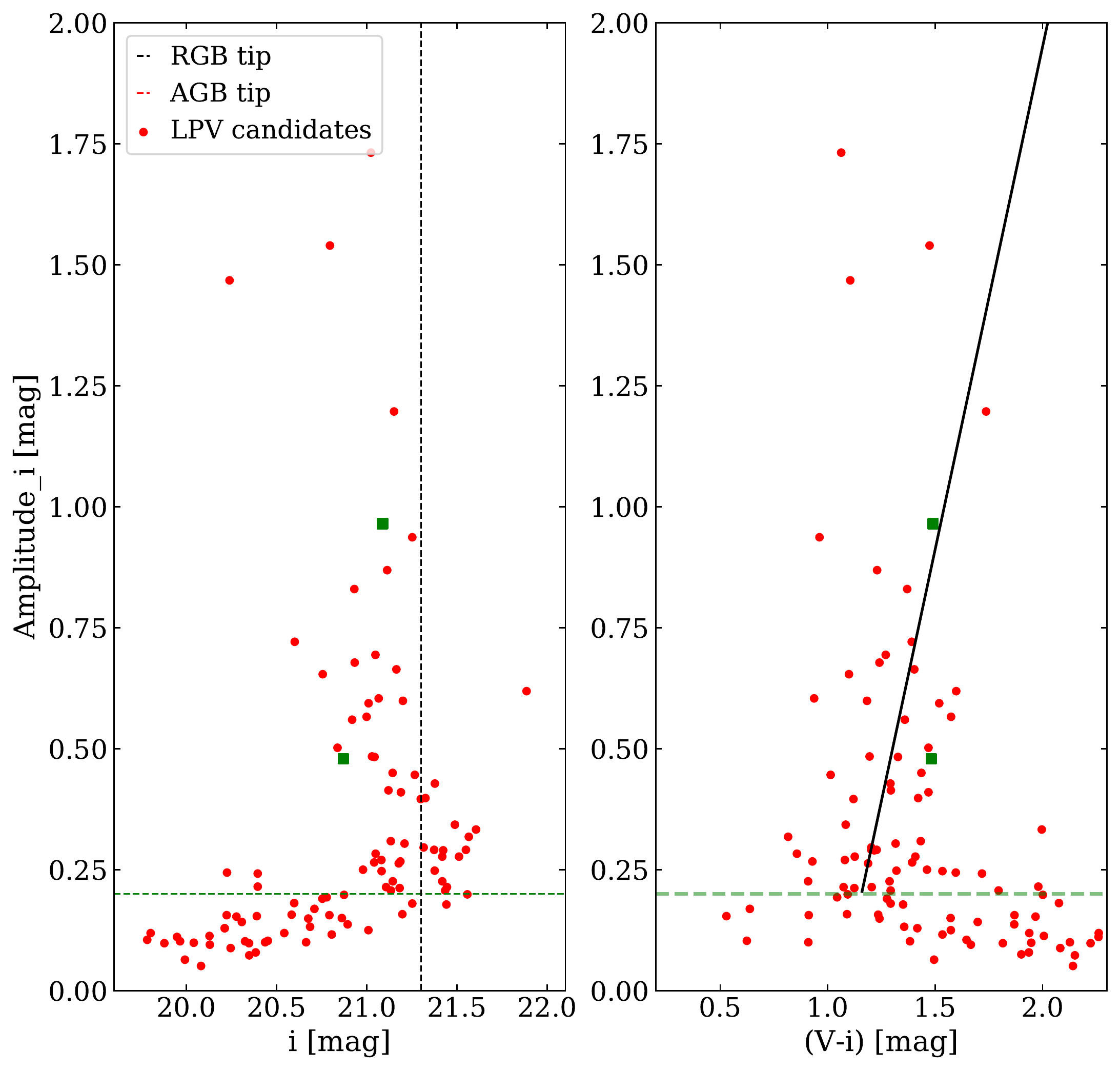}
	\caption{Estimated amplitude of variability $A_i$ vs.\ $i$-band magnitude (left panel) and $(V-i)$ color (right panel). The green dashed horizontal lines separate variables with $A_i<0.2$ mag. The Light-curves of the green points are displayed in Figure~\ref{fig:example of variable}. The RGB tip is indicated by the black dashed line in the left panel (see Section~\ref{Color-Magnitude Diagram})  and the solid black line shows the least-squares fit in the right panel. There is a high amplitude source with $A_i=3.78$ and color $(V-i)=1.64$ that it falls outside the boundaries of the graph.}
	\label{fig:amplitude}
	\end{figure}

Figure~\ref{fig:example of variable} shows the light-curve of a non-variable star (in the top panel), and two light-curves of candidate LPV stars labelled with green points in Figure~\ref{fig:amplitude} are presented in the middle and bottom panels. The amplitude of the light-curve can be estimated by assuming a sinusoidal shape. The root-mean-square of a sine wave with unit amplitude is 0.701, so the amplitude of a light-curve with a standard deviation of $\sigma$ can be calculated from $A=2\sigma/0.701$. We might not detect the faint or large-amplitude variables in some epochs when they drop below our detection threshold. In this case, we will underestimate their $i$-band variability.

The $i$-band amplitude of candidate LPV stars vs.\ their $i$-band magnitude is shown in the left panel of Figure~\ref{fig:amplitude}, and the amplitude vs.\ their $(V-i)$ color is presented in the right panel. A distinct feature of the left panel is that LPV amplitudes increase with decreasing brightness until $A_i\sim1.8$ mag \citep{2020svos.conf..383N}. In the right panel, the weighted average of colors for three different bins: 0.2 mag $<A_i<$ 0.5 mag,  0.5 mag $<A_i<$ 1 mag and $A_i>$ 1 mag mag are $1.21 \pm 0.03$ mag , $1.33 \pm 0.05$ mag and $1.61 \pm 0.12$ mag, respectively. Thus, it shows that LPV stars with larger amplitudes generally become redder.  Also, the solid black line represented in the right panel shows this trend by the weighted least-squares fitting for points with $A_i>0.2$. These trends seen in And\,VII are like those seen before in And\,I \citep{2020ApJ...894..135S}. As shown in Figure~\ref{fig:amplitude}, a number of candidate variable stars have $A_i<0.2$ mag and we cannot be certain about the nature of their variability. Therefore, we only consider high-amplitude variable stars as candidate LPV stars.

\begin{figure}
	\includegraphics[width=\columnwidth,height=1.03\columnwidth]{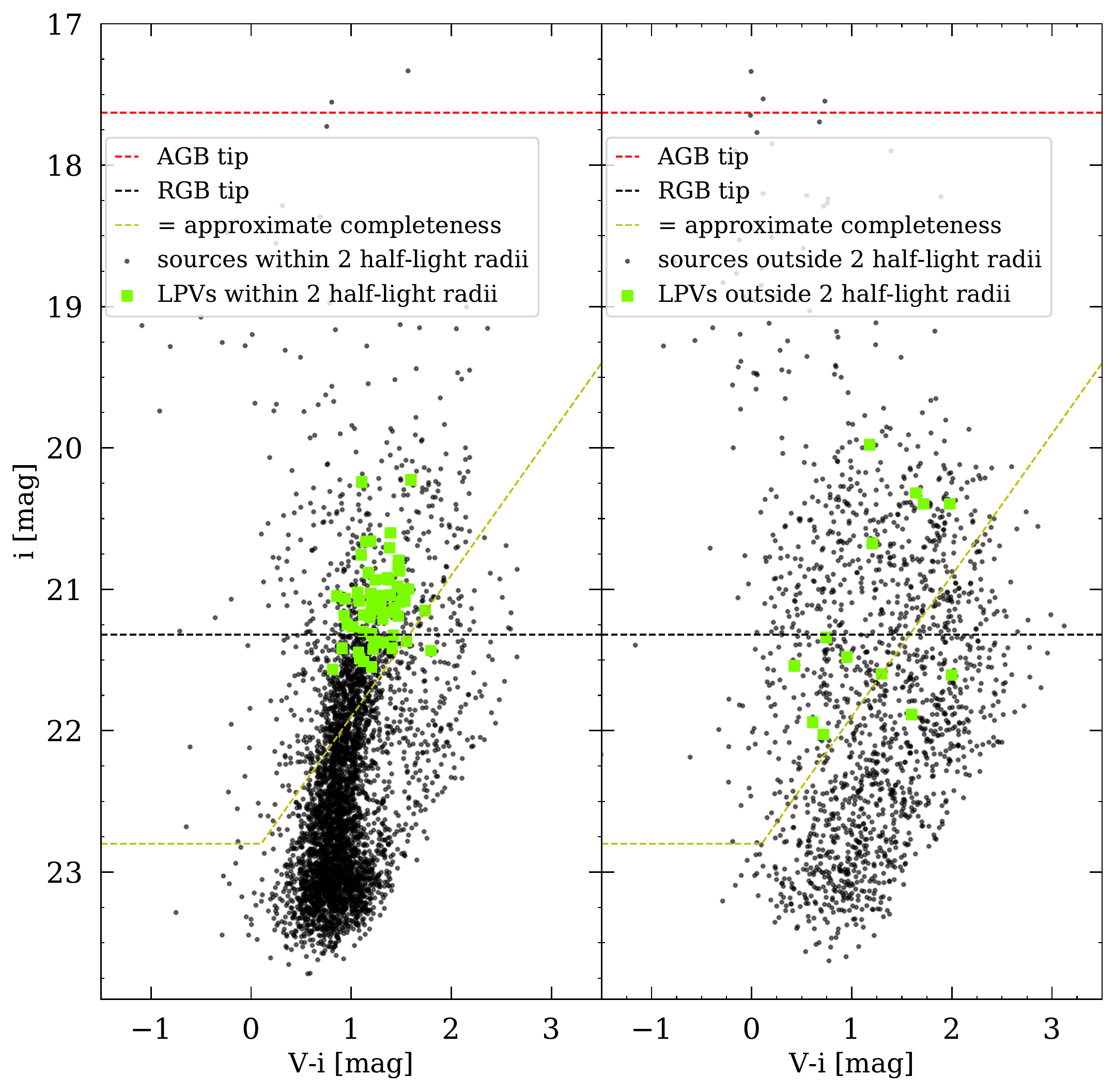}
	\caption{The CMD of And\,VII, showing our identified candidate LPV stars, in green, for two regions with similar areas: within two half-light radii (left panel) and outside two half-light radii (right panel). The latter shows mainly the contamination from foreground (and background) sources.}
	\label{fig:bothcmd}
\end{figure}

In the entire field of CCD 4, 62 candidate LPV stars have been identified, of which 43 are inside $r_{\frac{1}{2}}$ from the center of And\,VII. The density of candidate LPV stars inside $r_{\frac{1}{2}}$ is 1.1 arcmin$^{-2}$, which is much larger than the density of 0.03 arcmin$^{-2}$ outside this radius; hence, we can trust that most detected candidate LPV stars truly belong to And\,VII. Figure~\ref{fig:lpvs} shows the positions of the 55 LPV candidates within $2r_{\frac{1}{2}}$ in red and 7 out of the remaining 62 LPV candidates in blue, covering a similar area on the sky. Additionally, Figure~\ref{fig:bothcmd} shows the CMDs of these two regions (left and right panels for inside and outside $r_{\frac{1}{2}}$) with candidate LPV stars in green and other sources in black. By comparing the two CMDs we can estimate the total contamination from background galaxies and active galactic nuclei (AGN), and foreground stars that are below the limit of completeness of Gaia. Hence, within $2r_{\frac{1}{2}}$ of the center of And\,VII, $\sim46$\% of all stars and only 16\% of candidate LPV stars could be contaminants. The lack of a curved RGB outside $r_{\frac{1}{2}}$ (right panel of Figure~\ref{fig:bothcmd}) supports the notion that contamination from M\,31 is insignificant; hence, the Milky Way remains the major source of contamination. 

\begin{figure}
	\includegraphics[width=\columnwidth,height=1.\columnwidth]{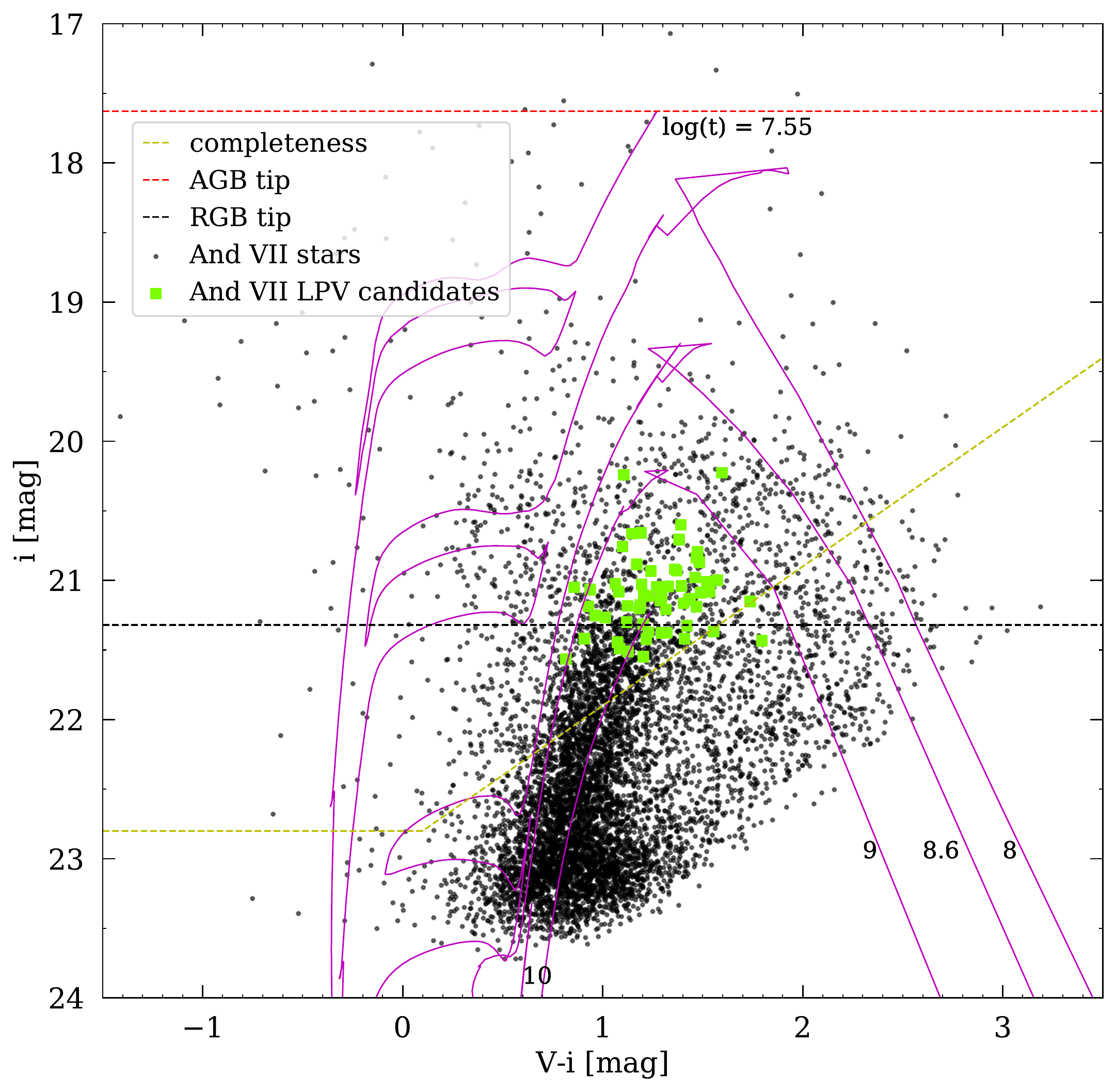}
	\caption{CMD of And\,VII with candidate LPV stars labeled in green colour. The CMD is overlain by the Padova stellar evolution models, and the AGB tip and RGB tip are illustrated by the red and black dashed lines, respectively. The 50\% completeness limit of our photometry is indicated with a yellow dashed line.}
	\label{fig:cmd1}
	%
	\includegraphics[width=\columnwidth,height=1.2\columnwidth]{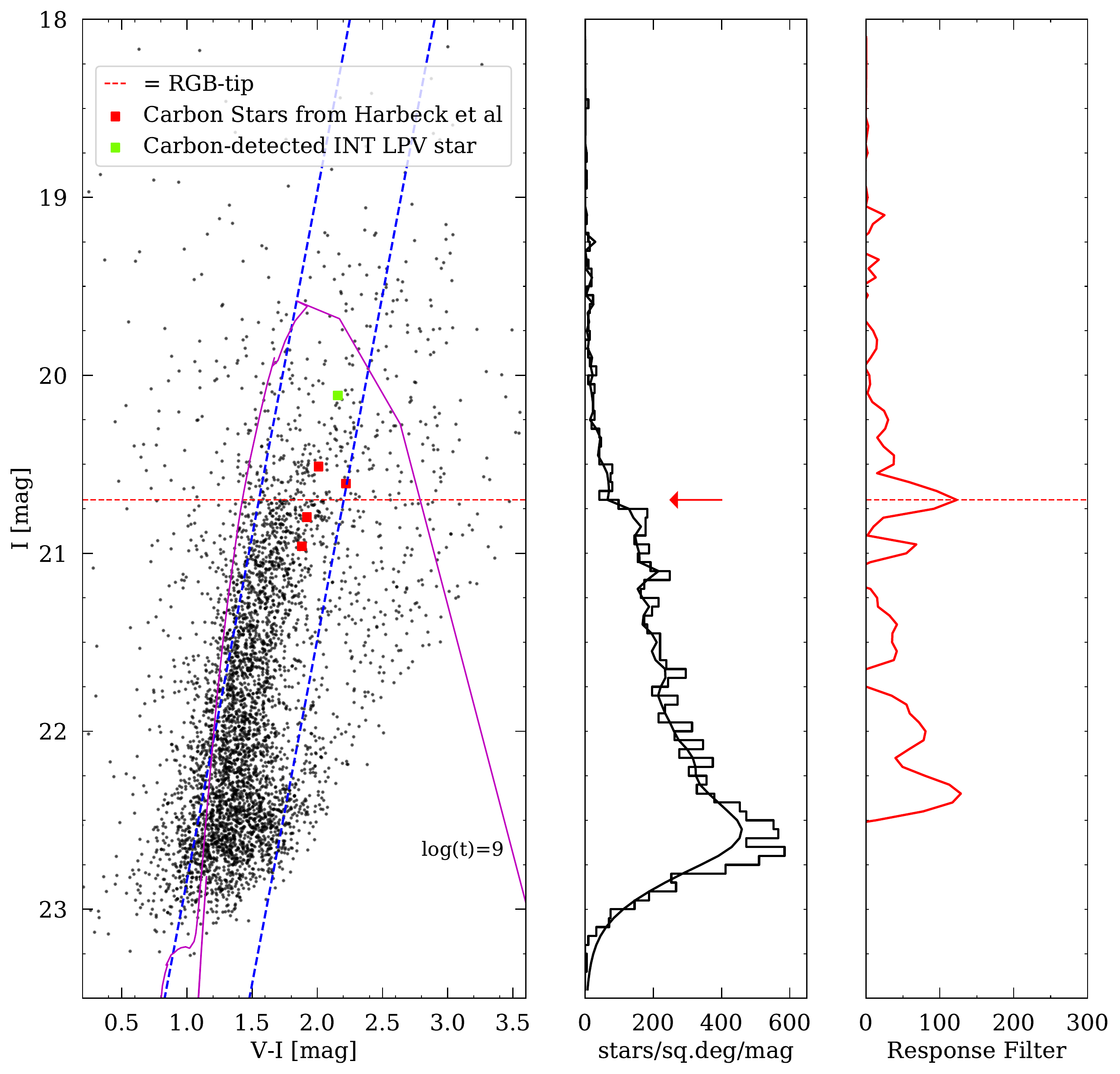}
	\caption{And\,VII CMD (left panel), luminosity function and LPD (middle), and response filter (right). The RGB tip is measured at $I=20.7$ mag and marked on the CMD by a horizontal red dotted line and on the luminosity functions by an arrow. The left diagram shows the carbon stars from the \cite{2004AAS...205.9301H} survey that were detected in our INT survey. A 1-Gyr isochrone is shown for comparison.}
	\label{fig:trgb}
\end{figure}
\section{Color--magnitude diagram}
\label{Color-Magnitude Diagram}

Figure~\ref{fig:cmd1} presents the CMD of $\sim 10\,000$ stars restricted to the $2r_{\frac{1}{2}}$ ellipse of And\,VII with 55 candidate LPV stars in green. The overplotted Padova {\sc parsec/colibri} isochrones \citep{2012MNRAS.427..127B,2017ApJ...835...77M,2019MNRAS.485.5666P} were computed using the {\sc cmd} v3.3 interface \footnote{http://stev.oapd.inaf.it/cgi-bin/cmd\_3.3}, adopting a metallicity of $Z=0.0007$, and covering a range of masses between 0.15--120 M$_\odot$. This model is derived for over 50 different photometric systems and instruments, including SDSS and Landolt, and predicts the large-amplitude pulsation, the related mass-loss rate and the dust formation rate. Also, the 50\% completeness limit of our photometry is determined based on a simple simulation in section~\ref{sec:Evaluation of Photometry}. The CMD demonstrates an early-type galaxy, with a prominent RGB branch and no evidence for young blue stellar sequences, but with a small intermediate-age AGB population immediately above it, dominated by our candidate LPV stars.

To quantify the AGB population, we must now define the magnitudes of the RGB and AGB tips, which are determined by the brightest star belonging to the RGB and AGB. In the RGB case, this corresponds to the start of helium burning. However, the AGB tip in this color combination does not represent the end of AGB evolution, but the point where the color--luminosity relationship breaks down due to circumstellar dust production, resulting in the optical obscuration of the stars. The RGB tip can be used as a standard candle, since the RGB luminosity distribution shows a discontinuity. To obtain the RGB tip, the $I$-band is most suitable as it exhibits minimal dependence on age and chemical composition \citep{1993ApJ...417..553L}. Therefore, we transformed the Sloan $i$ magnitudes to the $I$-band of the Johnson--Cousins system using the transformations from Lupton (2005). RGB stars are selected as those inside $r_{\frac{1}{2}}$ if they are located between the two blue lines on the CMD of Figure~\ref{fig:trgb}. Then, the luminosity probability distribution (LPD) was derived from them. The LPD technique is appropriate because there are a sufficient number of RGB stars and the effect of Poisson noise is modest \citep{2004MNRAS.350..243M}.

We show both the binned luminosity histogram of $I$ magnitudes (with 0.05-mag bins) and also the LPD (curved plot) in the middle panel of Figure~\ref{fig:trgb}. The right panel presents the convolution of the LPD with the Sobel kernel $[-2,-1,0,1,2]$. The peak of this response is marked by the red dashed line, which we identify as the RGB tip in the $I$-band ($20.7\pm0.05$ mag), from which we derive a distance modulus for the system of $\mu=24.38\pm0.05$ mag with the adopted value of $M^{\rm TRGB}_I$ and extinction from \cite{2004MNRAS.350..243M}. This result is in accordance with earlier studies (cf.\ Sec.~\ref{introduction}). We also obtained a magnitude for the RGB tip in the $i$-band of $21.3\pm0.05$ mag based on transforming again to Sloan $i$, which is shown in Figures~\ref{fig:bothcmd} and \ref{fig:cmd1} as a black dashed line.

We used the brightest point of isochrones to determine the theoretical AGB tip in the  CMD \citep[cf.][]{2020ApJ...894..135S}. The $M_{\rm bol}$ for a Chandrasekhar core mass is $-7.1$ mag, based on the classical core-luminosity relation. This core luminosity is reached by stars with $\log(t/{\rm yr})\lesssim 7.55$ in the adopted Padova isochrones. Accordingly, the peak of the $\log(t/{\rm yr})\sim 7.55$ isochrone reaches $i=17.63$ mag for an adopted metallicity of $Z=0.0007$ and distance modulus $\mu=24.38$ mag, which is shown by the red dashed line on the CMD of Figure~\ref{fig:cmd1}. Besides, the Padova {\sc parsec/colibri} isochrones produce carbon stars in the range $\log(t/{\rm yr})=8.0$--9.4 and $M_{\rm init}=1.3$--4.9 M$_\odot$. The limit to carbon-star masses is set by the efficiency of third dredge-up, which influences both when a carbon star is formed and when hot-bottom burning prevents a star from becoming carbon-rich.

Studies of carbon AGB stars in M\,31 dSph companions have recently been performed to map the extent of intermediate-age populations. \cite{2004AAS...205.9301H} found that most carbon stars reside in And\,VII among all M\,31 dwarf galaxies, so they discussed the presence of a substantial intermediate-age stellar population in And\,VII. To detect red giant stars with enhanced carbon abundance, they used two narrowband filters centered on the TiO and CN features at 778 and 808 nm. Carbon stars were identified on the basis of CN--TiO $\geq 0.2$ and $V-I>1.8$ mag and also a luminosity condition that carbon stars obey $I>22$ mag. Hence, they found in And\,VII three genuine carbon stars whose bolometric luminosities are brighter than the RGB tip and two carbon stars with dimmer luminosities than the RGB tip.

The five carbon candidates from \cite{2004AAS...205.9301H} are detected in the INT survey and identified in the CMD of Figure~\ref{fig:trgb} by red points. One of them is shown in green, since we also detect it as a candidate LPV star. All of them exhibit $(V-i)>1.8$ mag in the INT survey, likely indicating reddening due to circumstellar dust. The presence of carbon stars with lower luminosity than the RGB tip can be related to mass transfer from an erstwhile carbon star companion, or they may be intrinsic carbon stars at low metallicity \citep{2016ApJ...828...15H}. These stars are observed in other metal-poor galaxies and globular clusters, e.g., $\omega$\,Centauri \citep{2007MNRAS.382.1353V,2011MNRAS.417...20M} or Sagittarius dwarf irregular galaxy \citep{2012MNRAS.427.2647M}. It is worth mentioning that the DUSTiNGS star/extreme-AGB star catalog \citep{2015ApJS..216...10B} did not report any variable AGB stars in And\,VII. The reason is likely that two epochs are insufficient to catch all variability.

\begin{figure*}
	\includegraphics[width=2.12\columnwidth]{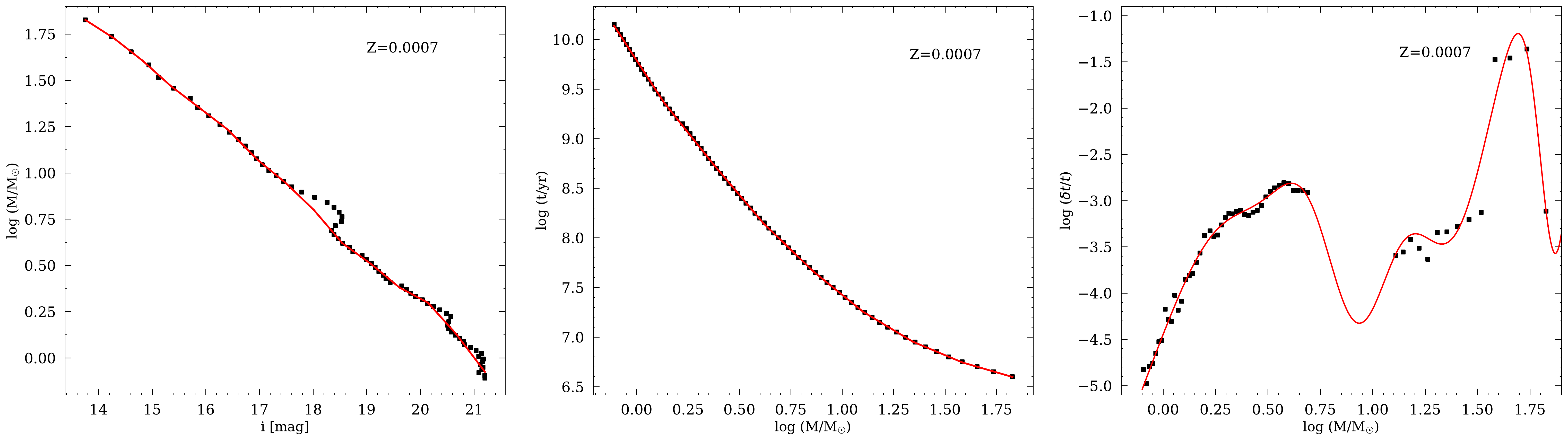}
	\caption{The relation between birth mass and $i$-band magnitude at the end points of stellar evolution for And\,VII, for a distance modulus of $\mu=24.38$ mag, interstellar attenuation $A_i=0.36$ mag, and metallicity $Z=0.0007$. Solid red lines are fits, in which the function is interpolated over the super-AGB phase ($0.6<\log(M/{\rm M}_\odot)<0.9$).}
	\label{fig:mass-mag}
\end{figure*}

\newpage
\section{Star formation history}
\label{sec:sfh}

To reconstruct the SFH, we compare our photometric data to the Padova models described in Section~\ref{Color-Magnitude Diagram}. LPV stars are assumed to be at their very ultimate point of evolution and to have reached their maximum luminosity. Thus, their brightness can be turned into their birth mass by applying theoretical evolutionary tracks. Hence, a mass--magnitude relationship was defined for the $i$ band by the brightest points on the Padova isochrones. This relationship is displayed in the left panel of Figure~\ref{fig:mass-mag} for isochrones with an age range of $6.5<\log(t/{\rm yr})<10.16$, with step size $\log(t/{\rm yr})=0.05$.

It is clear that there is a mostly linear relationship between mass and $i$-band magnitude. However, in the mass range $0.6<\log(M/{\rm M}_\odot)< 0.9$, an inversion appears as a result of the change in atmospheric composition, with oxygen-rich stars occurring during this mass range owing to hot-bottom burning, while the carbon stars below this mass range have brighter $i$-band magnitudes due to the different molecular opacities between TiO and C$\_2$/CN. Hence, the mass--magnitude relation was interpolated over this mass range and the best-fitting function on the other points was obtained with the {\sc iraf} task {\sc nf} -- displayed by red lines in Figure~\ref{fig:mass-mag}. The coefficients and intercepts for this mass--magnitude relation are recorded in Table~\ref{tab:mass-mag}. Accordingly, we can estimate the birth mass of LPV stars from this theoretical relation ($\log M/{\rm M}_\odot=a\times i+b$).

Furthermore, the relations between birth mass and age, and also between birth mass and pulsation duration are shown in the middle and right panels of Figure~\ref{fig:mass-mag}, respectively. The coefficients of the linear fits for mass--age and a set of five Gaussian fittings for mass--pulsation duration are listed in Tables~\ref{tab:mass-age} and \ref{tab:mass-pulse}, for a metallicity of $Z= 0.0007$ (relations for other metallicities used in this paper can be found in the Appendices: Figure~\ref{fig:mass-mag2,4} and Tables~\ref{tab:taba1},\ref{tab:taba2},\ref{tab:taba3}).

\subsection{Method of deriving the SFH}

The method used for deriving the SFH was developed by \cite{2011MNRAS.411..263J,2011MNRAS.414.3394J,2017MNRAS.464.2103J} and also applied and justified extensively by \cite{2014MNRAS.445.2214R,2017MNRAS.466.1764H,2019MNRAS.483.4751H}. The SFR is described as the rate at which gas is transformed into stars, $\xi$ (in M$_\odot$ yr$^{-1}$), as a function of time. The amount of stellar mass, $dM$, created in a time interval $dt$ is
\[dM(t) = \xi(t)\,dt\]

To describe the SFH, we used the LPV stars identified in And\,VII and their luminosity distribution function, $f(i)$, to reconstruct the birth mass function from the present stellar masses ($M$). Hence, the SFR as a function of look-back time ($t$) is
\begin{equation}
\xi(t) = \frac{f(i(M(t)))}{\delta(M(t))f_{IMF}(M(t))} 
\label{eq:SFH}
\end{equation}
where $\delta$ is the duration of the evolutionary phase during which LPV stars demonstrate strong radial pulsation and $f_{\rm IMF}$ is the initial mass function (IMF) that defines the proportional SFR of stars of different mass. Each of these functions are related to the stellar mass ($M$), and the mass of a pulsating star at the end of its evolution is directly associated with its age \citep{2011MNRAS.414.3394J}. With $f_{\rm IMF}$ defined by
\begin{equation}
f_{\rm IMF} = Am^{-\alpha},
\label{eq:alpha}
\end{equation}
where $A$ is the normalization constant and $\alpha$ depends on the mass range, following \cite{2001MNRAS.322..231K}:
\begin{equation}
\alpha= \begin{cases}
+0.3\pm0.7 ,& \text{if\space \space  \space} 0.02 <\frac{m}{{\rm M}_\odot}<0.08\\
+1.3\pm0.5 ,& \text{if\space \space  \space} 0.08<\frac{m}{{\rm M}_\odot}<0.5\\
+2.3\pm0.3 ,& \text{if\space \space  \space} 0.5<\frac{m}{{\rm M}_\odot}<200\\
\end{cases}
\end{equation}

There is convincing empirical evidence that LPV stars start to lose their mass and produce dust in this phase \cite[e.g.][]{2003A&A...401..347T, 2012A&A...545A..56Z, 2017A&A...597A..20O}. The bending of isochrones after their peak shows the evolution of dusty LPV stars that become dimmer and redder. Therefore, de-reddening correction for color and magnitude must apply to each LPV star that is under the effect of high dust column density. The magnitude correction in the $i$-band was obtained for the dusty LPV candidates on the CMD (Figure~\ref{fig:cmd1}) to return them to the expected magnitude (peak of isochrones). To this end, the slope of  $\log(t/{\rm yr})\simeq9$ isochrones are used for carbon stars and isochrones of  $\log(t/{\rm yr})=10$ or 8 are employed for M-stars. The average slopes of isochrones were derived as $a_{\rm oxygen}=2.04$ if $i\leq20.05$ mag and else $a_{\rm oxygen}=2.95$, and also, $a_{\rm carbon}=1.37$ if $i\leq20.87$ mag and otherwise $a_{\rm carbon}=3.65$. We applied the following correction equation for stars which have $(V-i)>1.4$ mag:
\begin{equation}
i_0 = i+a((V-i)_0-(V-i)),
\end{equation}
where the peaks of all isochrones are located at $(V-i)_0=1.16$ mag. We determined the type of star based on its mass: the mass range of $1.3<M/{\rm M}_\odot<4$ comprises carbon stars and below or above this range comprises oxygen stars. Among 55 candidate LPV stars, the presence of 16 carbon stars was estimated, out of which one carbon LPV star had previously been found by \cite{2004AAS...205.9301H} (Figure~\ref{fig:trgb}). The selected range for the mass of carbon stars is examined further in Section~\ref{sec:diffmasssfh}.

\begin{table}
	\caption{Fitted coefficients for the relation between birth mass and $i$-band magnitude, $\log M/{\rm M}_\odot=a\times i+b$ for a distance modulus of $\mu=24.38$ mag and a metallicity $Z=0.0007$.}
	\label{tab:mass-mag}
	\begin{tabular}{cccc}
		\hline\hline
		$a$    & $b$ & Validity Range \\
		\hline\hline
		$ -0.186\pm0.063 $ & $ 4.383\pm0.922 $ & $i\leq 14.285 $\\
		$ -0.227\pm0.064 $ & $ 4.968\pm0.967 $ & $ 14.285 <i\leq 14.817$\\
		$ -0.260\pm0.062 $ & $ 5.461\pm0.972 $ & $ 14.817<i\leq 15.349 $\\
		$ -0.221\pm0.055 $ & $ 4.863\pm0.879 $ & $ 15.349 <i\leq 15.882 $\\
		$ -0.223\pm0.047 $ & $ 4.889\pm0.786 $ & $ 15.822 <i\leq 16.414 $\\
		$ -0.296\pm0.042 $ & $ 6.085\pm0.714 $ & $ 16.414 <i\leq 16.946 $\\
		$ -0.239\pm0.042 $ & $ 5.124 \pm0.737 $ & $ 16.946 <i\leq 17.478 $\\
		$ -0.277\pm0.094 $ & $ 5.799\pm1.724 $ & $ 17.478 <i\leq 18.010$\\
		$ -0.338\pm0.091 $ & $ 6.892\pm1.699 $ & $ 18.010 <i\leq 18.542 $\\
		$ -0.210\pm0.037 $ & $ 4.520\pm0.710 $ & $ 18.542 <i\leq 19.075 $\\
		$ -0.240\pm0.037 $ & $ 5.081\pm0.727 $ & $ 19.075 <i\leq 19.607 $\\
		$ -0.155\pm0.039 $ & $  3.418\pm0.803 $ & $ 19.607 <i\leq 20.139 $\\
		$ -0.322\pm0.034 $ & $ -6.522\pm0.698 $ & $ 20.139 <i\leq 20.671 $\\
		$ -0.382\pm0.027 $ & $ -8.165\pm0.572 $ & $i> 20.671$ \\
		\hline\hline
	\end{tabular}
	\\
	\\
	\caption{The relation between age and birth mass, $\log t=a\log M/{\rm M}_\odot+b$ for a metallicity $Z=0.0007$. }
	\label{tab:mass-age}
	\begin{tabular}{cccc}
		\hline\hline
	 $a$    & $b$ & Validity Range  \\
		\hline	\hline
		$ -3.189\pm0.024 $ & $ 9.788\pm0.006 $ & $\log M\leq 0.133 $\\
		$ -2.594\pm0.022 $ & $ 9.709\pm0.011 $ & $ 0.133 <\log M\leq 0.375 $\\
		$ -2.441\pm0.023 $ & $ 9.652\pm0.017 $ & $ 0.375 <\log M\leq 0.617 $\\
		$ -2.002\pm0.025 $ & $ 9.382\pm0.025 $ & $ 0.617 <\log M\leq 0.859 $\\
		$ -1.680\pm0.028 $ & $ 9.105\pm0.034 $ & $ 0.859 <\log M\leq 1.101 $\\
		$ -1.248\pm0.032 $ & $ 8.629\pm0.047 $ & $ 1.101 <\log M\leq 1.343 $\\
		$ -0.867\pm0.037 $ & $ 8.118\pm0.064 $ & $ 1.343 <\log M\leq 1.585 $\\
		$ -0.601\pm0.045 $ & $ 7.696\pm0.088 $ & $\log M> 1.585 $ \\
		\hline\hline
	\end{tabular}
	\\
	\\
	
	\caption{Fits of the relation between relative pulsation duration ($\delta t/t$ where $t$ is the age and $\delta t$ is the pulsation duration) and birth mass, $\log(\delta t/t)=\Sigma_{n=1}^5a _n\exp\left[-(\log M[{\rm M}_\odot]-b_n)^2/c_n^2\right]$ for a metallicity $Z=0.0007$.}
	\label{tab:mass-pulse}
	\begin{tabular}{ p{1cm}p{2cm}p{2cm}p{2cm}}
		\hline
		\hline
		$i$&	$a$    & $b$ & $c$ \\
		\hline\hline
		1     &   $-82861.55$                 & $\phantom{-}1.033584$   & $0.345593 $  \\
		2     &   $\phantom{-}12.84194$ & $\phantom{-}0.769341 $   & $0.271284  $ \\
		3     &   $-8.071388$                     & $-0.878686$                    & $1.132971 $  \\
		4     &   $-3.383325 $                    & $\phantom{-}1.874786$   & $0.111842 $  \\
		5     &   $\phantom{-}82852.97$  & $\phantom{-}1.033613 $ & $1.033613 $ \\
		\hline\hline
	\end{tabular}
\end{table}

To derive the SFH, we followed the procedure for described by \cite{2011MNRAS.414.3394J}:
\begin{enumerate}
	\item Using the $i$-band carbon attenuation correction equation for each LPV's magnitude, we obtain the original luminosity. Then, if the mass is not between 1.1--4 M$_\odot$, we de-redden with oxygen correction.
	\item Using the corrected $i$-band magnitude and mass--magnitude relation in Table~\ref{tab:mass-mag}, the birth mass is determined.
	\item Using the age--mass relation in Table~\ref{tab:mass-age}, the age is determined.
	\item Using the pulsation duration--mass relation in Table~\ref{tab:mass-pulse}, the pulsation duration of long-period variability is determined.
	\item Applying equations~\ref{eq:SFH} and \ref{eq:alpha}, we calculate the SFR for selected age bins that have the same number of LPV stars.
	\item Using Poisson statistics, we calculate the statistical error for each bin \citep{2017MNRAS.466.1764H}.
\end{enumerate}

\begin{figure}	
	\includegraphics[width=\columnwidth,height=\columnwidth]{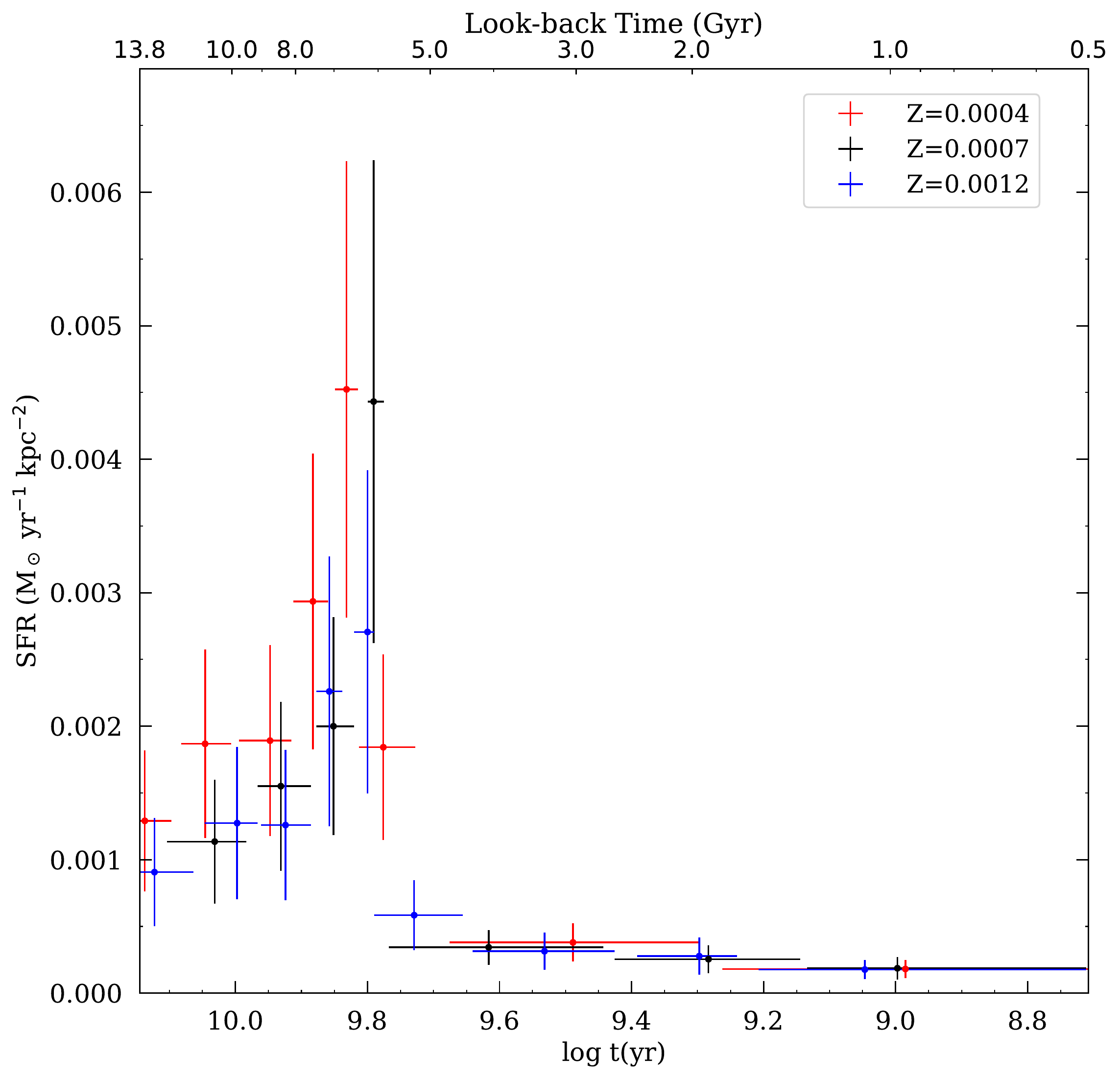}
	\caption{SFH of And\,VII within $2r_{\frac{1}{2}}$ for three assumed (constant) metallicities of $Z=0.0004$, 0.0007 and 0.0012 in red, black and blue color, respectively.}
	\label{fig:SFH,2rh}			
	\includegraphics[width=\columnwidth,height=\columnwidth]{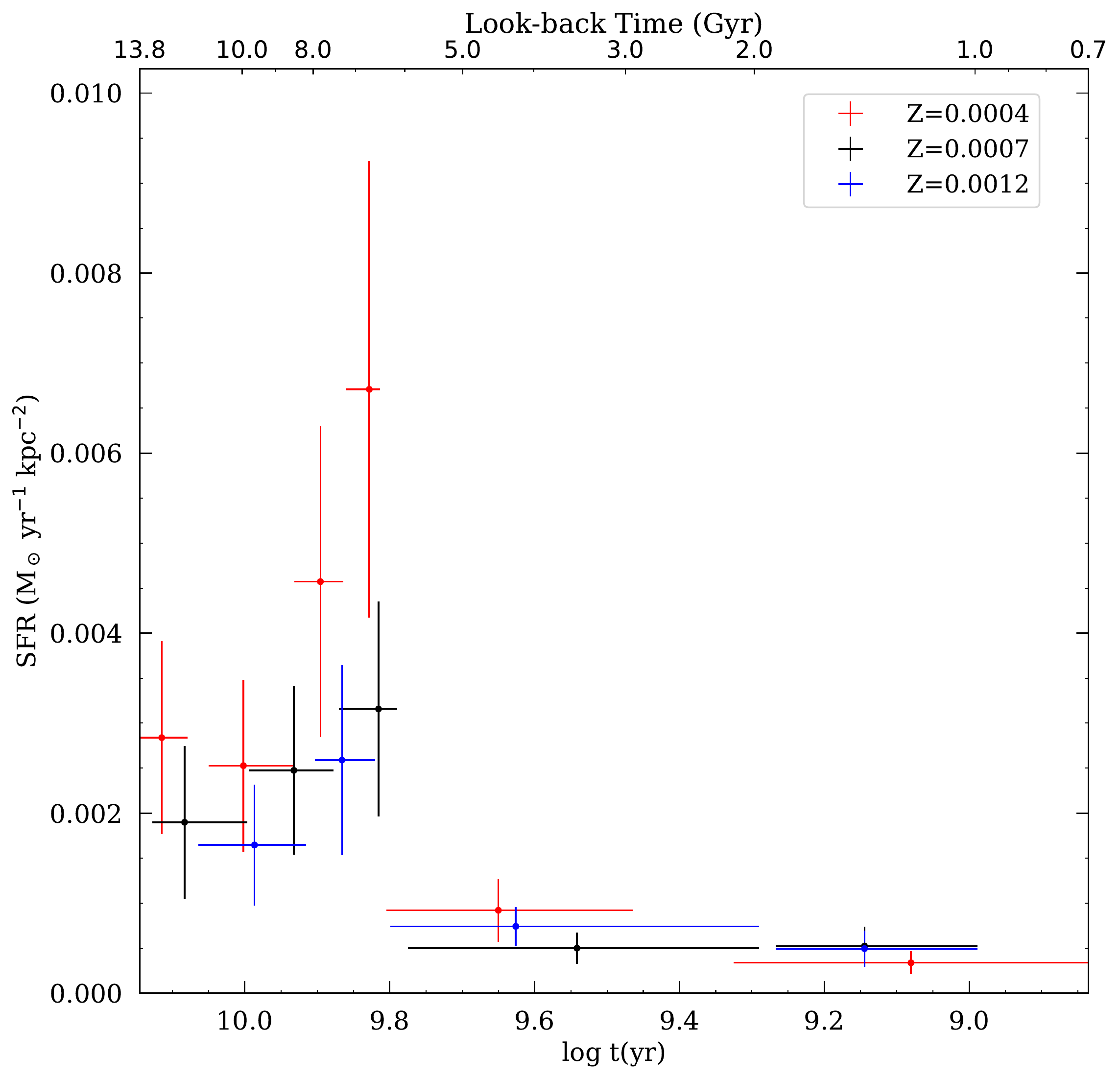}
	\caption{SFH of And\,VII within $r_{\frac{1}{2}}$ for three assumed (constant) metallicities of $Z=0.0004$, 0.0007 and 0.0012 in red, black and blue color, respectively.}
	\label{fig:SFH,rh}			
\end{figure}

\section{Results and discussion}
\label{sec:result}

The estimated SFH of And\,VII within $2r_{\frac{1}{2}}$ and $r_{\frac{1}{2}}$ of the galaxy's centre for different constant metallicities, over an interval from 13.8 Gyr ($\log t=10.14$) to $\sim 1$ Gyr ago ($\log t=9$) is presented in Figures~\ref{fig:SFH,2rh} and \ref{fig:SFH,rh}, respectively. The SFH is characterized by bins which have lengths indicated by horizontal error bars; the vertical error bars show the statistical errors on the SFR in each bin. As we expect for dSph galaxies, star formation continued up to a specific time (perhaps infalling time) and, after that, star formation appears to have stopped entirely. For And\,VII, star formation continued up until $\sim1$ Gyr ago and the SFH shows evidence of an intermediate-age stellar population.

Assuming $Z=0.0007$ within $2r_{\frac{1}{2}}$, the peak of star formation occurred $\sim6.2$ Gyr ago when the SFR reached a level of $0.0044\pm0.0018$ M$_\odot$ yr$^{-1}$ kpc$^{-2}$ on average over an interval of $\sim 0.2$ Gyr. This is at least twelve times as high as the next three epochs between 6 and 1 Gyr ago. Approximately 90\% of the total stellar mass ($13.9\times10^6$ M$_\odot$) in And\,VII was formed during the period of intense star formation from 13.8 to 6 Gyr ago. After the peak, the SFR dropped suddenly and continued at a rate of $\simeq 0.0003\pm0.0002$ M$_\odot$ yr$^{-1}$ kpc$^{-2}$ until one Gyr ago. However, in the central part ($r\leq r_{\frac{1}{2}}$) the SFR peaked at a lower rate, $0.003\pm0.001$ M$_\odot$ yr$^{-1}$ kpc$^{-2}$ at $\log t({\rm yr})\sim9.8$.

The total stellar mass ($M_{\rm tot}$) formed inside $r_{\frac{1}{2}}$ and $2r_{\frac{1}{2}}$ was estimated to be $(13.3\pm5.3)\times10^6$ M$_\odot$ and $(15.4\pm6.2)\times10^6$ M$_\odot$, respectively (Table~\ref{tab:mass}). They are in agreement with the values obtained by previous studies, such as \cite{2020AJ....159...46K}, who reported $M_{\rm tot}=(16.2\pm1.3)\times10^6$ M$_\odot$, and \cite{2012AJ....144....4M} who obtained $M_{\rm tot}=9.5\times10^6$ M$_\odot$ but $M_{\rm tot}=19.7\times10^6$ M$_\odot$ in \citeyear{2018ApJ...868...55M}. Furthermore, \cite{2014ApJ...789..147W} counted the total stellar mass formed within $0.19r_{\frac{1}{2}}$ as $(12.8\pm1.5)\times10^6$ M$_\odot$. However, their shallow photometry made their SFH in this galaxy less secure. They found And\,VII to have an ancient SFH, while the presence of metal-rich, high-[$\alpha$/Fe] stars, which were found by \cite{2020AJ....159...46K,2020ApJ...895...78W}, do not agree with an exclusively ancient SFH.

In order to follow back the SFH, it is necessary to take into account the variation of metallicity with age up to the quenching time. Since dSphs have been quenched early on, their chemical evolution did not significantly change at a later time \citep{2007IAUS..235...57S}. Metallicity of the ISM is a function of the look-back time of a galaxy, and increases with decreasing $\alpha$-element abundance. However, And\,VII in all metallicity ranges exhibits a constant value of [$\alpha$/Fe] $\simeq +0.3$ \citep{2014ApJ...790...73V} or shows a shallow negative slope in [$\alpha$/Fe]--[Fe/H] with a mean value of [$\alpha$/Fe] $\simeq +0.18\pm0.03$ \citep{2020AJ....159...46K,2020ApJ...895...78W}. By taking into account the $\alpha$-element abundance for the metallicity of [Fe/H] $\simeq -1.24\pm0.003$ dex, we considered a metal-rich environment of $Z=0.0012$ for the young population. In contrast, for the old population $Z=0.0004$ was adopted as stars with [Fe/H] $\in [-2.5,-1.5]$ have been found in And\,VII. Hence, Figures~\ref{fig:SFH,rh} and \ref{fig:SFH,2rh} show how the SFR changes for different assumptions of the metallicity: the SFRs for $t>8$ Gyr at $Z=0.0004$ and for $t<5$ Gyr at $Z=0.0012$ are larger than the corresponding SFR at $Z=0.0007$.

\begin{figure}
	\includegraphics[width=\columnwidth]{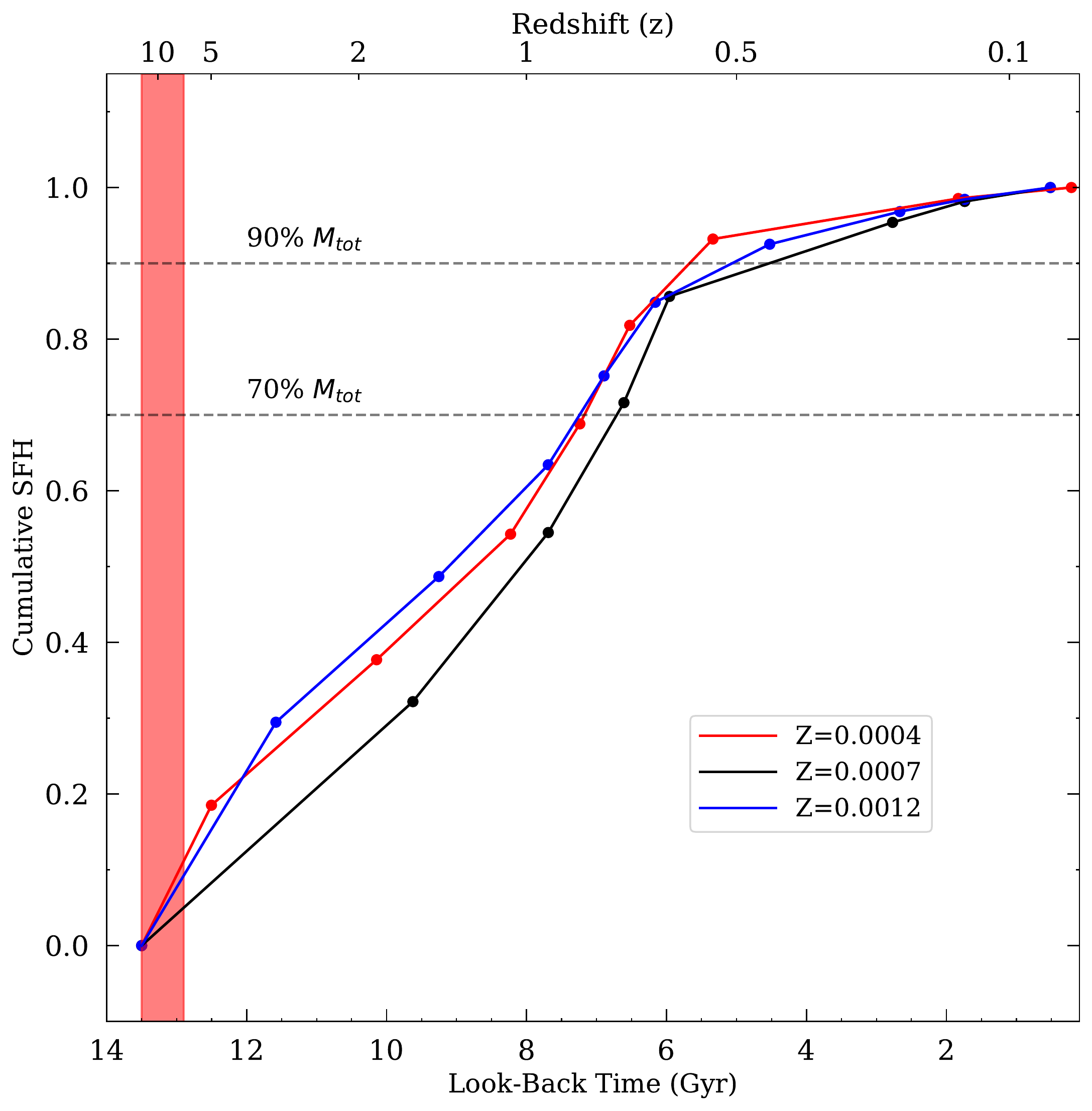}
	\caption{Cumulative SFH of And\,VII within $2r_{\frac{1}{2}}$ for adopted metallicities of $Z=0.0007$ and 0.0004. The red band indicates the epoch of reionization \citep[$z\sim6$--14, 12.9--13.5 Gyr ago; ][]{2006ARA&A..44..415F}.}
	\label{fig:cumsfhboth}
\end{figure}

\begin{table}
	\caption{Total stellar mass for different metallicities and regions.}
	\label{tab:mass}
	\begin{tabular}{ p{2cm}p{3cm}cp{2cm}}
		\hline \hline
		$Z$   & $M_{\rm tot}$ ( $10^6$  M$_\odot$) & Regions \\
		\hline	\hline
		0.0012 & $9.5\pm4.2$   & r$_h$   \\
		0.0012 & $17.7\pm7.9$  & 2 r$_h$   \\
		0.0007 & $13.3\pm5.3$  & r$_h$   \\
		0.0007 & $15.4\pm6.2$  & 2 r$_h$ \\
		0.0004 & $20.6\pm8.4$  & r$_h$ \\  
		0.0004 & $26.8\pm10.1$ & 2 r$_h$ \\   
		\hline\hline
	\end{tabular}
\end{table}

\subsection{Cumulative SFH and quenching time}

The cumulative SFH is built up of stellar mass over cosmological time:
\begin{equation}
Cumulative\ SFH\pm\delta=\Sigma\frac{SFR\pm error_{\rm SFR}}{M_{\rm cumulative}}\times\Delta t,
\end{equation}
where $\Delta t$ specifies the length of each temporal bin and $error_{\rm SFR}$ is that bin's uncertainty in SFR. The cumulative SFH of And\,VII, assuming the two metalicities $Z=0.0007$ and 0.0004, is shown in Figure~\ref{fig:cumsfhboth}. From that, we can calculate the amount of stellar mass at any given time. Dwarf spheroidal galaxies hold very little gas and the star formation in them has been extinguished many Gyr ago. The quenching time (hereafter $\tau_{90}$) for each of them is when 90\% of the total stellar mass had been formed. By interpolating the cumulative SFH of And\,VII, $\tau_{90}$ is estimated about 5 Gyr ago ($\log t=9.7$) with the present metal-rich condition ($\tau_{90}=4.97^{+0.95}_{-2.30}$ Gyr ago with $Z=0.0007$ and $\tau_{90}=5.06^{+1.09}_{-0.69}$ Gyr ago with $Z=0.0012$), but if we assume that the galaxy initially formed iron-poor stars, the quenching time may have happened as early as $5.77^{+0.73}_{-1.03}$ Gyr ago ($\log t=9.76$) with $Z=0.0004$.

\cite{2015ApJ...804..136W} investigated the cumulative SFH of 38 dwarf galaxies in the Local Group and concluded that the mass of dwarfs plays an important role in the quenching time. Although they did not report the quenching time of And\,VII in their study, we can compare And\,VII to other M\,31 dSphs (And\,I, II, III, VI) with the same mass. The $\tau_{90}$ of these galaxies were found from 7.4--5.2 Gyr ago ($\log(t/{\rm yr})=9.87$--9.72) by \cite{2015ApJ...804..136W} and the quenching time of And\,VII, with the largest mass, is later than most, if not all of them. Our results confirm the observed relation that lower mass dwarfs typically cease stellar mass assembly earlier than higher mass galaxies \citep{2019MNRAS.489.4574G,2015ApJ...804..136W}. We note that these five galaxies are located within the M\,31 virial radius where the quenching time is thus essentially independent of the current proximity to M\,31 \citep{2015ApJ...804..136W}.

Although most M\,31 dSph satellites include old stellar populations with ages $\geq 10$ Gyr, \cite{2004AAS...205.9301H} declared that And\,VII could possess an intermediate-age population due to the evidence of carbon stars in this dwarf. Accordingly, they mentioned that star formation in And\,VII may have been active as recently as 3--5 Gyr ago. Also, they remarked based on the model of \cite{2003MNRAS.338..572M} that there is no evidence of a notably younger intermediate-age ($<3$ Gyr) stellar population in And\,VII. Moreover, \cite{2015ApJ...807...49W} applied the {\sc elvis} suite to estimate the distribution of infall times for satellites with $M_{\rm star}=10^{3-9}$ M$_\odot$ at $z=0$. They realized that infall into the Milky Way/M\,31 halo typically happened 5--8 Gyr ago at $z=0.5$--1.  Assuming that environmental quenching is related to infall time, And\,VII first fell into the M\,31 halo about 5--8 Gyr ago, leading to suppression of its star formation at 5 Gyr ago.

The red band in Figure~\ref{fig:cumsfhboth} indicates the epoch of reionization \citep[$z\sim6$--14, 12.9--13.5 Gyr ago; ][]{2006ARA&A..44..415F}. The best fossil candidates at reionization are low-mass dwarfs ($M_{\rm star}\lesssim 10^6$ M$_\odot$), where at least 70\% of their stellar mass had been formed by the time of cosmic reionization. The dashed line in Figure~\ref{fig:cumsfhboth} shows that $\tau_{70}$ (i.e.\ the epoch by which 70\% of the total stellar mass had formed) was in place by 7 Gyr ago at $Z=0.0007$ or 7.7 Gyr ago at $Z=0.0004$. Therefore, it has not been presumed that quenching of And\,VII (with a large mass of $\sim 10^7$ M$_\odot$) was affected by reionization. Most Local Group dwarfs did not form the bulk of their stellar mass during reionization \citep{2014ApJ...789..148W} and probably environment has a stronger impact on the quenching of their star formation \citep{2015ApJ...804..136W}.

Also, $\tau_{70}$ can define the duration and efficiency of star formation in dSphs. The relation between dark halo structure and SFH of several M\,31 dwarf galaxies were investigated based on comparisons between the density profile of dark halo ($\rho(b_{\rm halo})$) and $\tau_{70}$ by \cite{2015IAUGA..2254520H}. Their aim was to figure out whether dark halo properties depend on the SFH of the stellar component. For this purpose, they adopted $\tau_{70}$ values from \cite{2014ApJ...789..147W} who reported $\tau_{70}\simeq 12.8$ Gyr for And\,VII, which caused this dwarf to be classified as a galaxy with a rapid SFH that holds a dense and concentrated dark halo. However, \cite{2014ApJ...789..147W} were not satisfied with the SFH result of And\,VII from {\it Hubble} space Telescope WFPC2 data in their own research since the CMD of this galaxy was too shallow to show any age-sensitive features. Fortunately, in this study, we could calculate the exact value of And\,VII's $\tau_{70}$ being between 7.0--7.7 Gyr, thus the SFH of this galaxy has not happened as rapidly as \cite{2014ApJ...789..147W} predicted. Moreover, this value is more consistent with $\tau_{70}$ of other similar M\,31 dSphs \citep{2014ApJ...789..148W}. Also, $\rho(b_{\rm halo})$ of And\,VII is more similar to M\,31 dwarf galaxies with $\tau_{70}$ between 7--10 Gyr \citep{2015IAUGA..2254520H}.


\begin{figure}
	\includegraphics[width=\columnwidth,height=0.9\columnwidth]{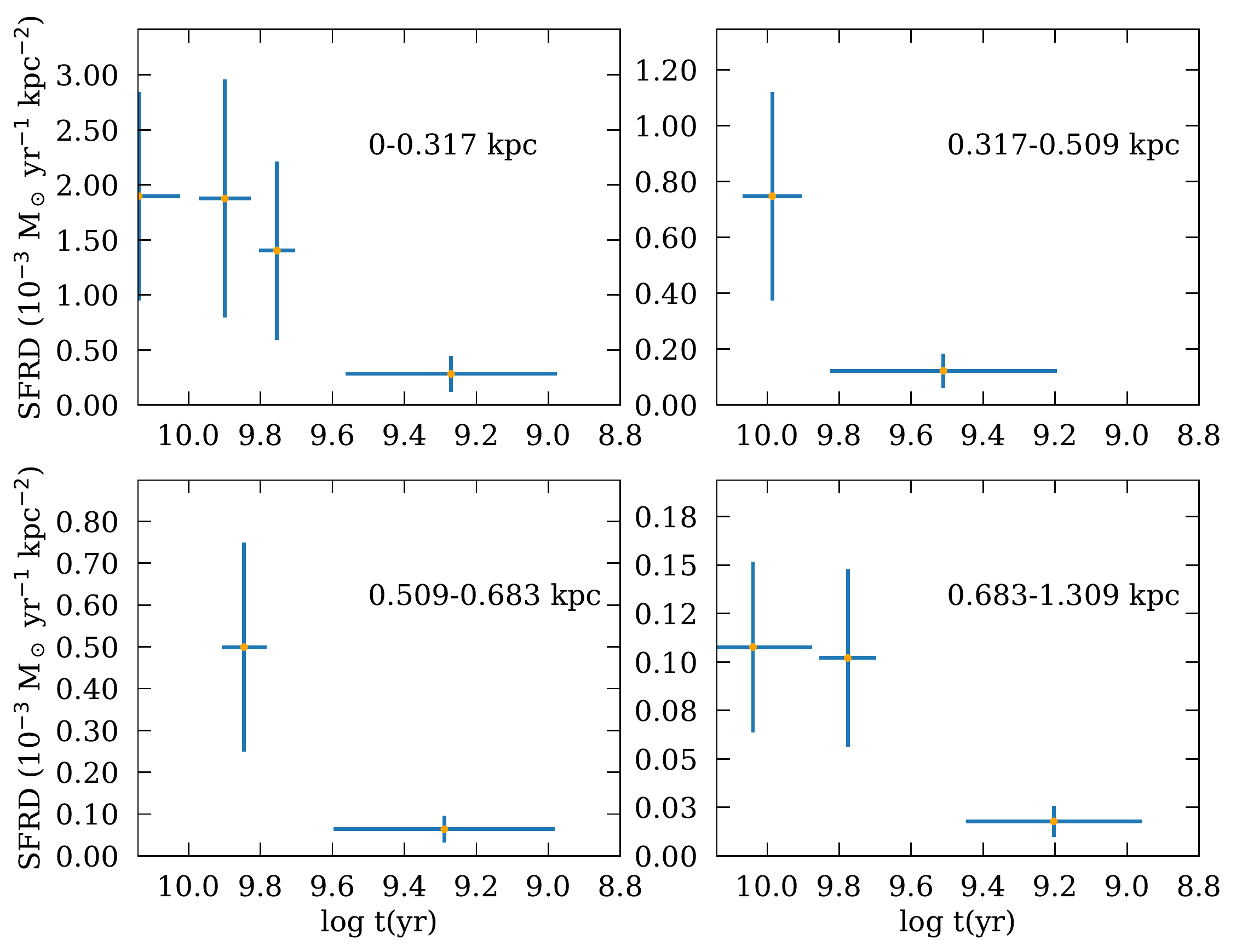}
	\caption{SFHs of And\,VII at constant $Z=0.0007$ in bins with equal numbers of LPV stars at galactocentric radii as indicated.}
	\label{fig:galactocentric}
\end{figure}

\subsection{Galactocentric radial gradient of SFH}

The galactocentric radial gradient of SFH traces the compactness of star formation across cosmological time. The SFHs of And\,VII across different radius bins with an equal number of stars are displayed in Figure~\ref{fig:galactocentric} at $Z=0.0007$. It is clear that there is not a distinct trend in with radius, and star formation at any age has occurred at all distances. This result was also concluded by \cite{2014ApJ...790...73V} based on the $\alpha$-element abundance gradient. Since And\,VII does not exhibit any evidence of an $[\alpha/Fe]$ radial gradient, they concluded that the star-formation timescales were not strongly dependent on the radial distance from the center of the galaxy. Figure~\ref{fig:galactocentric} demonstrates that the peak of SFH occurred at all radii before 6 Gyr ago and its rate has been decreasing with increasing radial distance. The peak SFR declines by a factor $\sim 15$ between the innermost and outermost regions ($\delta r\sim1.25$ kpc). The most recent star formation ($\log t/{\rm yr}<9.6$) is more noticeable in the inner regions ($<0.5$ kpc), suggesting star formation may have been quenched outside--in.


\begin{figure*}	
	\includegraphics[width=2\columnwidth,height=0.94\columnwidth]{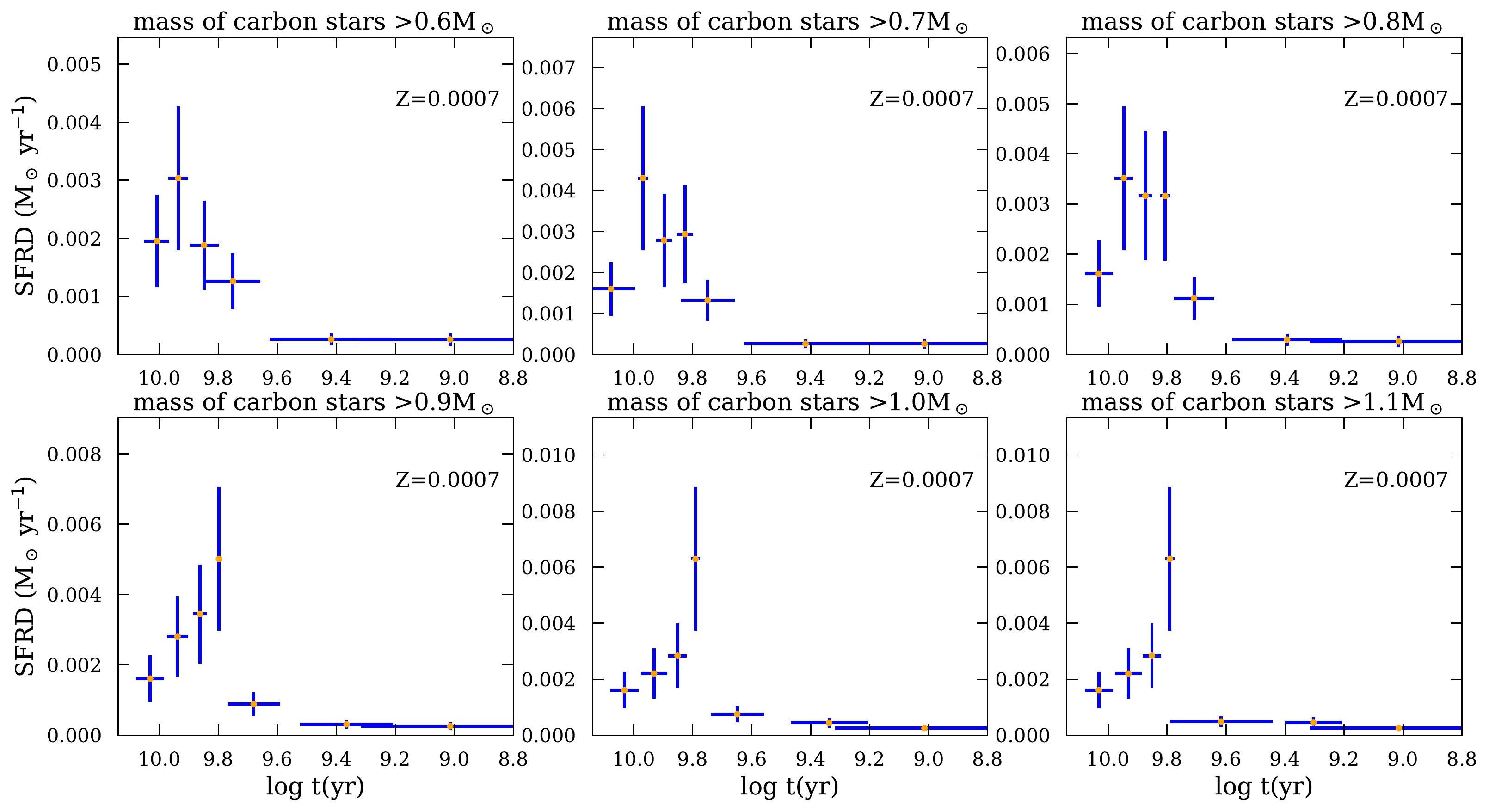}
	\caption{Remeasuring of And\,VII's SFH with different mass ranges for carbon stars. From the top-left to the down-right panels, the minimum of mass range for C-stars is decreasing (from $0.6\rightarrow1.1$ M$_\odot$).}
	\label{fig:diffmassSFH}
\end{figure*}

\subsection{Minimum birth mass of carbon stars}
\label{sec:diffmasssfh}

Since the majority of our candidate LPV stars do not have observational information to distinguish carbon stars from oxygen-rich (M-type) stars, we had to select a mass range for carbon stars based on theory (Padova model). However, the masses of carbon stars that were found by \cite{2004AAS...205.9301H} in And\,VII are obtained between 0.6--1.7 using our modeling. At a metallicity of $Z=0.004$ (typical of the Magellanic Clouds), the AGB stars with a minimum birth mass $\sim1.1$ M$_\odot$ were estimated to be a carbon star \citep[e.g.][]{1981A&A....94..175R,1993A&A...267..410G,2016MNRAS.460.4230D}. Also, \citep{2002PASA...19..515K,2014MNRAS.445..347K} predicted that at low metallicity, the efficiency of dredge-up is higher, and low-mass AGB stars can more easily turn into carbon stars.

Accordingly, at $Z=0.0007$ or lower, the mass range of carbon stars can be larger and carbon stars might be produced from a minimum birth mass $<1.1$ M$_\odot$. Hence, we remeasured And\,VII's SFH based on carbon stars that arise from stars with a minimum of solar and subsolar initial mass (from 0.6--1.1 M$_\odot$). The lower end of this range implies that the results are obtained using exclusively carbonaceous dust correction. In Figure~\ref{fig:diffmassSFH}, we show the SFH for the different mass range of carbon stars. Considering the subsolar carbon stars, the SFH would have experienced an exponential decrease, but without them, star formation in And\,VII suddenly decreased after the peak. In Table~\ref{tab:diffmasssfh}, the differences between total masses and quenching times are small. It appears that a different mass range for carbon stars does not substantively alter the inferred time of quenching but only has an impact on the manner in which it was quenching.

\begin{table}
	\caption{Total stellar mass and quenching time for SFHs with different minimal initial mass for carbon stars.}
	\label{tab:diffmasssfh}
	\begin{tabular}{ cp{3cm}cp{2cm}cp{5cm}}
		\hline \hline
		birth mass M$_\odot$  & $M_{\rm tot}$ ($10^7$ M$_\odot$) & $\tau_{90}$ (Gyr)\\
		\hline	\hline
		$0.6$&$1.66\pm0.67$& $5.2^{+1.3}_{-0.4}$   \\
		$0.7$&$1.71\pm0.70$  &$5.2^{+1.0}_{-0.7}$\\
		$0.8$&$ 1.55 \pm 0.64 $&$5.1^{+1.0}_{-0.7}$ \\  
		$0.9$&$1.50 \pm0.62 $ &$4.9^{+1.1}_{-0.9}$ \\   
		$1.0$&$1.46 \pm 0.60$ & $4.8^{+1.1}_{-1.1}$\\ 
		$1.1$&$ 1.44\pm 0.59$ & $4.9^{+1.3}_{-1.7}$\\ 
		\hline\hline
	\end{tabular}
\end{table}

\section{Conclusions}
\label{sec:conclusions}

We have presented a photometric catalogue in the $i$ and $V$ filters from new data obtained with INT/WFC of the And\,VII dwarf spheroidal galaxy. About 10\,000 stars, including 55 candidate LPV stars, were detected within twice the half-light radius, determined as $3\rlap{.}^\prime8$ in this study. The AGB and RGB tips of the galaxy are identified. By combining the numbers and brightnesses of detected LPV stars with stellar isochrones, the SFH of And\,VII was reconstructed. The main epoch of star formation in And\,VII happened $\approx 6.2$ Gyr ago and no evidence of star formation more recent than 400 Myr ago can be found in this galaxy. About 70\% of And\,VII's total stellar mass may have been assembled by 7 Gyr ago and 90\% was in place by 5 Gyr ago, when this dSph galaxy was quenched. The quenching of this galaxy occurred after reionization, and likely it was related to environmental impacts. Thus, And\,VII is dominated by an old (7--14 Gyr) stellar population. But also it includes metal-rich intermediate-age (5--7 Gyr) stars, which appears confirmed by the known presence of five carbon stars in this dwarf \citep{2004AAS...205.9301H}. Consequently, And\,VII contains a young population of similar age to the stellar halo of M\,31. The total mass of stars that was formed during the galaxy's evolution is $(15.4\pm6.2)\times10^6$ M$_\odot$ within an elliptical radius of $2r_{\frac{1}{2}}\simeq 1.5$ kpc from the center.

\section*{Acknowledgement}
The observing time for this survey was primarily provided by the Iranian National Observatory (INO), complemented by UK-PATT allocation of time to programmes I/2016B/09 and I/2017B/04 (PI: J.\ van Loon). We thank the INO and the School of Astronomy (IPM) for the financial support of this project. Authors are grateful to Peter Stetson for sharing his photometry routines. We thank James Bamber, Philip Short, Lucia Su\'arez-Andr\'es and Rosa Clavero for their help with the observations.



\appendix
The relations between birth mass and parametres such as $i$-band magnitude, age and pulsation duration for metallicites of $Z= 0.0004$ and  0.0012 are presented in the Figure~\ref{fig:mass-mag2,4} ,Tables~\ref{tab:taba1},\ref{tab:taba2},\ref{tab:taba3}).
	
\begin{figure*}
	\includegraphics[width=\columnwidth]{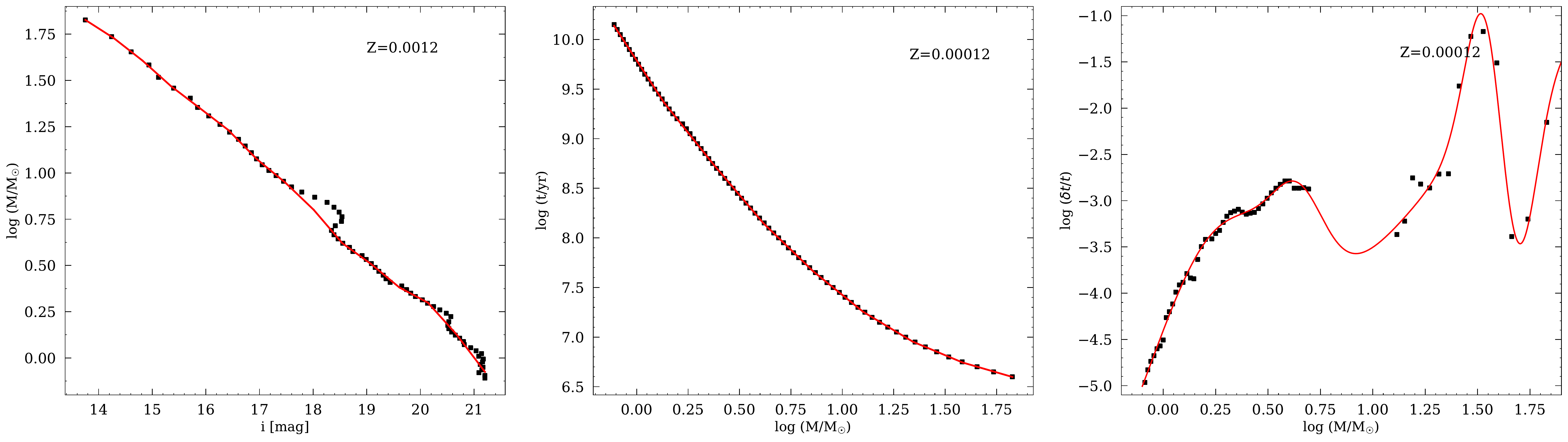}
	\includegraphics[width=\columnwidth]{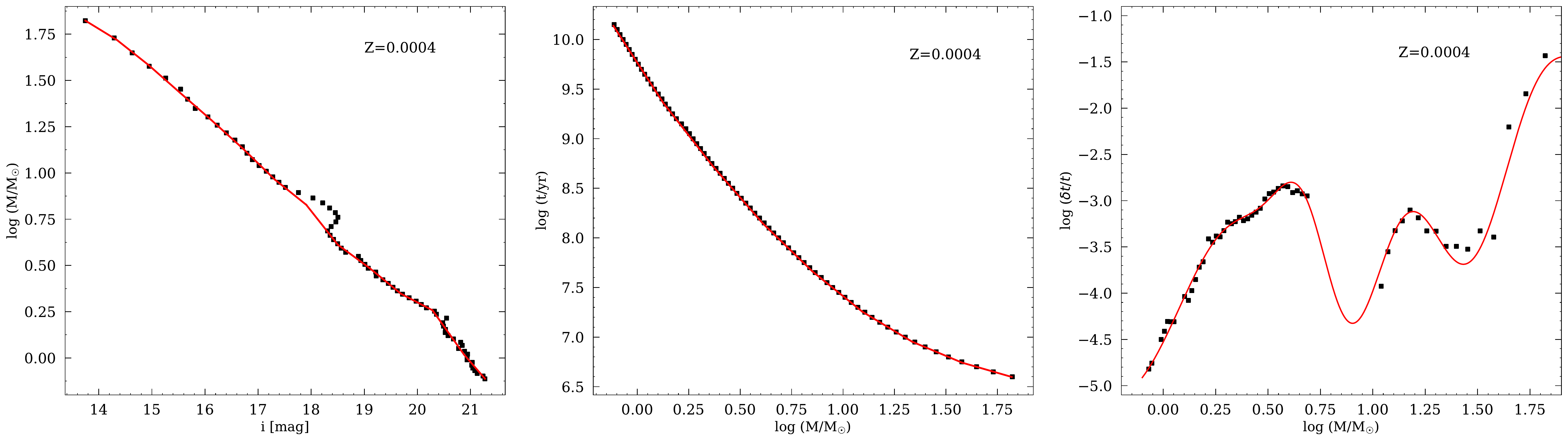}	
	\caption{The relations between birth mass and parameters such as $i$-band magnitude, age and pulsation duration are shown from left to right panels, respectively (at a distance modulus of $\mu=24.38$ mag for  $Z= 0.0012$ in the top row and for $Z=0.0004$ in the below row).}
	\label{fig:mass-mag2,4}
\end{figure*}

\begin{table}
	\caption{The fitting equations of the relation between birth mass and $i$-band magnitude, $\log M/{\rm M}_\odot=ai+b$ for a distance modulus $\mu=24.38$ mag.}
				\label{tab:taba1}
	\begin{tabular}{ccr}
		\hline\hline
		$a$              & $b$              & validity range    \vspace{0.01 in}\\
		\hline
		\multicolumn{3}{c}{$Z=0.0012$} \\
		\hline	
		$ -0.186\pm0.063 $ & $ 4.383\pm0.922 $ & $i\leq 14.285 $\\
		$ -0.227\pm0.064 $ & $ 4.968\pm0.967 $ & $ 14.285 <i\leq 14.817$\\
		$ -0.260\pm0.062 $ & $ 5.461\pm0.972 $ & $ 14.817<i\leq 15.349 $\\
		$ -0.221\pm0.055 $ & $ 4.863\pm0.879 $ & $ 15.349 <i\leq 15.882 $\\
		$ -0.223\pm0.047 $ & $ 4.889\pm0.786 $ & $ 15.822 <i\leq 16.414 $\\
		$ -0.296\pm0.042 $ & $ 6.085\pm0.714 $ & $ 16.414 <i\leq 16.946 $\\
		$ -0.239\pm0.042 $ & $ 5.124 \pm0.737 $ & $ 16.946 <i\leq 17.478 $\\
		$ -0.277\pm0.094 $ & $ 5.799\pm1.724 $ & $ 17.478 <i\leq 18.010$\\
		$ -0.338\pm0.091 $ & $ 6.892\pm1.699 $ & $ 18.010 <i\leq 18.542 $\\
		$ -0.210\pm0.037 $ & $ 4.520\pm0.710 $ & $ 18.542 <i\leq 19.075 $\\
		$ -0.240\pm0.037 $ & $ 5.081\pm0.727 $ & $ 19.075 <i\leq 19.607 $\\
		$ -0.155\pm0.039 $ & $  3.418\pm0.803 $ & $ 19.607 <i\leq 20.139 $\\
		$ -0.322\pm0.034 $ & $ -6.522\pm0.698 $ & $ 20.139 <i\leq 20.671 $\\
		$ -0.382\pm0.027 $ & $ -8.165\pm0.572 $ & $i> 20.671$ \\
		\hline
	\end{tabular}
	\begin{tabular}{ccr}
		\hline\hline
		$a$              & $b$              & validity range    \vspace{0.01 in}\\
		\hline
		\multicolumn{3}{c}{$Z=0.0004$} \\
		\hline
		$ -0.176\pm0.052 $ & $ 4.242\pm0.754 $ & $i\leq 14.342 $\\
		$ -0.228\pm0.049 $ & $ 4.983\pm0.741 $ & $ 14.342 <i\leq 14.936$\\
		$ -0.253\pm0.050 $ & $ 5.356\pm0.790 $ & $ 14.936 <i\leq 15.530 $\\
		$ -0.251\pm0.047 $ & $ 5.325\pm0.762 $ & $ 15.530 <i\leq 16.125 $\\
		$ -0.259\pm0.035 $ & $ 5.465\pm0.600 $ & $16.125 <i\leq 16.719 $\\
		$ -0.279\pm0.033 $ & $ 5.792\pm0.589 $ & $ 16.719 <i\leq 17.313 $\\
		$ -0.231\pm0.120 $ & $ 4.956\pm2.205 $ & $ 17.313 <i\leq 17.907 $\\
		$ -0.363\pm0.118 $ & $ 7.334\pm2.186 $ & $ 17.907 <i\leq 18.501$\\
		$ -0.211\pm0.029 $ & $ 4.513\pm0.563 $ & $ 18.501<i\leq 19.095 $\\
		$ -0.235\pm0.027 $ & $ 4.971\pm0.545 $ & $ 19.095 <i\leq 19.689 $\\
		$ -0.150\pm0.027 $ & $ 3.309\pm0.546 $ & $19.689<i\leq 20.284$\\
		$ -0.403\pm0.022 $ & $ 8.425\pm0.462 $ & $ 20.284 <i\leq 20.878 $\\
		$ -0.220\pm0.029 $ & $ -4.795\pm0.596 $ & $i>20.878 $\\
		\hline
	\end{tabular}
\end{table}

\begin{table}
	\caption{Relation between age and birth mass, $\log t=a\log M+b$.}
				\label{tab:taba2}
	\begin{tabular}{ccr}
		\hline \hline
		$a$              & $b$             & validity range\vspace{0.01 in}\\
		\hline
		
		\multicolumn{3}{c}{$Z=0.0012$} \\
		\hline
		$ -3.189\pm0.024 $ & $ 9.788\pm0.006 $ & $\log M\leq 0.133 $\\
		$ -2.594\pm0.022 $ & $ 9.709\pm0.011 $ & $ 0.133 <\log M\leq 0.375 $\\
		$ -2.441\pm0.023 $ & $ 9.652\pm0.017 $ & $ 0.375 <\log M\leq 0.617 $\\
		$ -2.002\pm0.025 $ & $ 9.382\pm0.025 $ & $ 0.617 <\log M\leq 0.859 $\\
		$ -1.680\pm0.028 $ & $ 9.105\pm0.034 $ & $ 0.859 <\log M\leq 1.101 $\\
		$ -1.248\pm0.032 $ & $ 8.629\pm0.047 $ & $ 1.101 <\log M\leq 1.343 $\\
		$ -0.867\pm0.037 $ & $ 8.118\pm0.064 $ & $ 1.343 <\log M\leq 1.585 $\\
		$ -0.601\pm0.045 $ & $ 7.696\pm0.088 $ & $\log M> 1.585 $ \\
		\hline
	\end{tabular}
	\begin{tabular}{ccr}
		\hline \hline
		$a$              & $b$             & validity range\vspace{0.01 in}\\
		\hline
		\multicolumn{3}{c}{$Z=0.0004$} \\
		\hline
		$ -3.201\pm0.022 $ & $ 9.776\pm0.005 $ & $log M\leq 0.129 $\\
		$ -2.616\pm0.020 $ & $ 9.700\pm0.010 $ & $ 0.129 <log M\leq 0.371 $\\
		$ -2.404\pm0.021 $ & $ 9.622\pm0.016 $ & $ 0.371 <log M\leq 0.613 $\\
		$ -2.006\pm0.023 $ & $ 9.377\pm0.023 $ & $ 0.613 <log M\leq 0.855 $\\
		$ -1.685\pm0.026 $ & $ 9.103\pm0.031 $ & $ 0.855 <log M\leq 1.097 $\\
		$ -1.249\pm0.029 $ & $ 8.625\pm0.043 $ & $ 1.097 <log M\leq 1.339 $\\
		$ -0.869\pm0.035 $ & $ 8.116\pm0.059 $ & $ 1.339 <log M\leq 1.581 $\\
		$ -0.595\pm0.042 $ & $ 7.683\pm0.082 $ & $log M> 1.581 $\\
		\hline
	\end{tabular}
\end{table}

\begin{table}
	\caption{Fits to the relation between relative pulsation duration ($\delta t/t$ where $t$ is the age and $\delta t$ is the pulsation duration) and birth mass, $\log(\delta t/t)=\Sigma_{i=1}^5a _i\exp\left[-(\log M[{\rm M}_\odot]-b_i)^2/c_i^2\right]$.}
				\label{tab:taba3}
	\begin{tabular}{ p{1cm}p{2cm}p{2cm}p{2cm}}
		\hline
		\hline
		i&	a    & b &c \\
		\hline
		\multicolumn{4}{c}{$Z=0.0012$} \vspace{0.1 in}     \\
		1     &   $0.7591$                      &  $0.6472$                   & $0.1710$  \\
		2     &   $2.6567$                     &  $0.2388$    & $0.4794$   \\
		3     &   --$27.203 $ &  $1.61884$     & $0.14442$  \\
		4     &   --$6.8127$ & --$0.4083$    & $ 1.81376$  \\
		5     &   $27.3586 $                    &  $1.60982$       & $0.14418$  \vspace{0.1 in} \\		
		\hline		
		\multicolumn{4}{c}{$Z=0.0004$} \vspace{0.1 in}  \\
		
		1     &   $-32.369$                      &  $\phantom{-}6.8349 $  & $5.4629$  \\
		2     &   $\phantom{-}0.4006$  &  $\phantom{-}0.2669$  & $0.0976$ \\
		3     &   $\phantom{-}35.049 $  &  $\phantom{-}1.3629$   & $0.8449$  \\
		4     &   $-1.5269$                      &  $\phantom{-}1.5129$   & $0.1107$  \\
		5     &   $-27.069$                     &  $\phantom{-}1.2729$   & $0.6624$  \vspace{0.1 in} \\					
		\hline
	\end{tabular}
\end{table}

\end{document}